\documentclass[fleqn,usenatbib]{mnras}

\usepackage{newtxtext,newtxmath}

\usepackage[T1]{fontenc}

\DeclareRobustCommand{\VAN}[3]{#2}
\let\VANthebibliography\thebibliography
\def\thebibliography{\DeclareRobustCommand{\VAN}[3]{##3}\VANthebibliography}

\usepackage{graphicx}	
\usepackage{amsmath}	
\usepackage{scalerel}
\usepackage{tikz}
\usetikzlibrary{svg.path}

\definecolor{orcidlogocol}{HTML}{A6CE39}
\tikzset{
  orcidlogo/.pic={
    \fill[orcidlogocol] svg{M256,128c0,70.7-57.3,128-128,128C57.3,256,0,198.7,0,128C0,57.3,57.3,0,128,0C198.7,0,256,57.3,256,128z};
    \fill[white] svg{M86.3,186.2H70.9V79.1h15.4v48.4V186.2z}
                 svg{M108.9,79.1h41.6c39.6,0,57,28.3,57,53.6c0,27.5-21.5,53.6-56.8,53.6h-41.8V79.1z M124.3,172.4h24.5c34.9,0,42.9-26.5,42.9-39.7c0-21.5-13.7-39.7-43.7-39.7h-23.7V172.4z}
                 svg{M88.7,56.8c0,5.5-4.5,10.1-10.1,10.1c-5.6,0-10.1-4.6-10.1-10.1c0-5.6,4.5-10.1,10.1-10.1C84.2,46.7,88.7,51.3,88.7,56.8z};
  }
}

\newcommand\orcidicon[1]{\href{https://orcid.org/#1}{\mbox{\scalerel*{
\begin{tikzpicture}[xscale=0.1,yscale=-0.1,transform shape]
\pic{orcidlogo};
\end{tikzpicture}
}{|}}}}

\usepackage{hyperref} 

\makeatletter
\newcommand*{\rom}[1]{\expandafter\@slowromancap\romannumeral #1@}
\makeatother

\newcommand{\gcm}{g\,cm$^{-3}$}	

\newcommand{\tess}{{\it TESS}}

\newcommand{\harps}{{HARPS}}

\newcommand{\kms}{km\,s$^{-1}$}
\newcommand{\ms}{m\,s$^{-1}$}

\newcommand{\rstar}{\mbox{$R_{\star}$}}

\newcommand{\msun}{\mbox{$M_{\odot}$}}
\newcommand{\rsun}{\mbox{$R_{\odot}$}}
\newcommand{\rearth}{R$_{\oplus}$}
\newcommand{\mearth}{M$_{\oplus}$}

\newcommand{\vsini}{$v\sin{i}$}
\newcommand{\teff}{$T_{\rm eff}$}
\newcommand{\feh}{\mbox{$\rm [Fe/H]$}}
\newcommand{\ymg}{\mbox{$\rm [Y/Mg]$}}
\newcommand{\logg}{$\log g$}

\newcommand{\Tlogg}{ $ 4.46 \pm 0.05 $ }
\newcommand{\TFeH}{ $ 0.14 \pm 0.02 $ }
\newcommand{\TTeff}{ $ 5732 \pm 50 $ }
\newcommand{\Tmstartorres}{ $ 0.989 \pm 0.01 $ }
\newcommand{\Trstartorres}{ $ 0.998 \pm 0.025 $ }

\newcommand{\Tradiso}{ $ 1.002 \pm 0.028 $ }
\newcommand{\Tmassiso}{ $ 1.004^{+0.042}_{-0.047} $ }

\newcommand{\Tmstar}{ $ 0.997 \pm 0.06 $ }
\newcommand{\Trstar}{ $ 0.968 \pm 0.018 $ }

\newcommand{\Ttzerozero}{ $ 1570.101^{+0.004}_{-0.005} $ }
\newcommand{\Ttzeroone}{ $ 1798.17 \pm 0.19 $ }
\newcommand{\TKzero}{ $ 2.15 \pm 0.28 $ }
\newcommand{\TlogKzero}{ $ 0.77 \pm 0.13 $ }
\newcommand{\TKone}{ $ 3.58 \pm 0.37 $ }
\newcommand{\TlogKone}{ $ 1.27 \pm 0.1 $ }

\newcommand{\Trvlogerrcontrzero}{ $ -0.5^{+0.7}_{-1.7} $ }

\newcommand{\Trvlogerrcontrone}{ $ -12.84^{+0.53}_{-0.74} $ }

\newcommand{\Trvlogerrcontrtwo}{ $ 0.1 \pm 1.1 $ }

\newcommand{\Tphotlogerrcontr}{ $ -8.0^{+1.7}_{-2.6} $ }
\newcommand{\Tmeans}{ $ -0.0^{+0.001}_{-0.002} $ }
\newcommand{\Tmeanfwhm}{ $ 7286.9 \pm 1.1 $ }

\newcommand{\Tlogamprv}{ $ 3.66^{+0.35}_{-0.34} $ }

\newcommand{\Tlogamps}{ $ -9.64^{+0.37}_{-0.33} $ }

\newcommand{\Tlogampfwhm}{ $ 3.77 \pm 0.37 $ }

\newcommand{\TlogQzero}{ $ -1.1^{+1.6}_{-3.6} $ }
\newcommand{\TdeltaQ}{ $ 1.7^{+1.8}_{-1.4} $ }

\newcommand{\TPzero}{ $ 2.541^{+0.0005}_{-0.001} $ }
\newcommand{\TPone}{ $ 6.744^{+0.008}_{-0.009} $ }
\newcommand{\Tecczero}{ $ 0.093^{+0.079}_{-0.064} $ }
\newcommand{\Teccone}{ $ 0.045^{+0.079}_{-0.038} $ }
\newcommand{\Tomegazero}{ $ -0.47 \pm 0.68 $ }
\newcommand{\Tomegaone}{ $ -0.1 \pm 1.2 $ }
\newcommand{\TMpzero}{ $ 4.55 \pm 0.62 $ }
\newcommand{\TMpone}{ $ 10.5 \pm 1.2 $ }
\newcommand{\Tlogror}{ $ -4.009 \pm 0.063 $ }
\newcommand{\Tror}{ $ 0.018 \pm 0.001 $ }
\newcommand{\Trpl}{ $ 2.05 \pm 0.12 $ }
\newcommand{\Tb}{ $ 0.31 \pm 0.22 $ }
\newcommand{\Trhopgcmthree}{ $ 2.90^{+0.75}_{-0.59} $ }
\newcommand{\Tperiod}{ $ 20.8 \pm 1.2 $ }
\newcommand{\Tlogperiod}{ $ 3.035 \pm 0.06 $ }

\newcommand{\TphotSzero}{ $ 0.012 \pm 0.003 $ }
\newcommand{\Tphotwzero}{ $ 3.73^{+0.48}_{-0.44} $ }
\newcommand{\Tphotmean}{ $ 0.011^{+0.031}_{-0.035} $ }

\newcommand{\Tsmazero}{ $ 0.035 \pm 0.001 $ }
\newcommand{\Tsmaone}{ $ 0.068^{+0.001}_{-0.002} $ }
\newcommand{\TSinzero}{ $ 1001.0 \pm 40.0 $ }
\newcommand{\TSinone}{ $ 272.0 \pm 11.0 $ }
\newcommand{\TTsurfpzero}{ $ 1371.0 \pm 14.0 $ }
\newcommand{\TTsurfpone}{ $ 990.0 \pm 10.0 $ }
\newcommand{\Ttdurzero}{ $ 0.099^{+0.005}_{-0.007} $ }

\newcommand{\Trvtrendzero}{ $ 0.027 \pm 0.01 $ }
\newcommand{\Trvtrendone}{ $ -0.24 \pm 0.7 $ }
\newcommand{\Tustartesszero}{ $ 0.37 \pm 0.1 $ }
\newcommand{\Tustartessone}{ $ 0.221^{+0.093}_{-0.096} $ }

\newcommand{\Tdepth}{410\,ppm}

\newcommand{\TTstar}{TOI-755}
\newcommand{\TTplanet}{TOI-755.01}
\newcommand{\Tstar}{HD\,110113}
\newcommand{\Tstarage}{4.0\,$\pm$\,0.5 Gyr}
\newcommand{\Tplanet}{HD\,110113\,b}
\newcommand{\Tplanetc}{HD\,110113\,c}
\newcommand{\TGAIAid}{6133384959942131968}

\newcommand{\TdeltaBIC}{$16.32$}

\title[A hot mini-Neptune orbiting \Tstar{}]{A hot mini-Neptune in the radius valley orbiting solar analogue \Tstar{}}

\author[H.P. Osborn et al.]{\parbox{\textwidth}{H.P. Osborn\textsuperscript{\hyperlink{affil_1}{1},\hyperlink{affil_2}{2},\orcidicon{0000-0002-4047-4724}}\thanks{E-mail: hugh.osborn@space.unibe.ch}, 
D.J.~Armstrong\textsuperscript{\hyperlink{affil_3}{3},\hyperlink{affil_4}{4},\orcidicon{0000-0002-5080-4117}}, 
V.~Adibekyan\textsuperscript{\hyperlink{affil_5}{5},\orcidicon{0000-0002-0601-6199}}, 
K.A.~Collins\textsuperscript{\hyperlink{affil_6}{6},\orcidicon{0000-0001-6588-9574}}, 
E.~Delgado-Mena\textsuperscript{\hyperlink{affil_5}{5},\orcidicon{0000-0003-4434-2195}}, 
S.B.~Howell\textsuperscript{\hyperlink{affil_7}{7},\orcidicon{0000-0002-2532-2853}}, 
C.~Hellier\textsuperscript{\hyperlink{affil_8}{8},\orcidicon{0000-0002-3439-1439}}, 
G.W.~King\textsuperscript{\hyperlink{affil_3}{3},\hyperlink{affil_4}{4},\orcidicon{0000-0002-3641-6636}}, 
J.~Lillo-Box\textsuperscript{\hyperlink{affil_9}{9},\orcidicon{0000-0003-3742-1987}}, 
L.D.~Nielsen\textsuperscript{\hyperlink{affil_10}{10},\orcidicon{0000-0002-5254-2499}}, 
J.F.~Otegi\textsuperscript{\hyperlink{affil_10}{10}}, 
N.C.~Santos\textsuperscript{\hyperlink{affil_5}{5},\hyperlink{affil_11}{11},\orcidicon{0000-0003-4422-2919}}, 
C.~Ziegler\textsuperscript{\hyperlink{affil_12}{12}}, 
D.R.~Anderson\textsuperscript{\hyperlink{affil_3}{3},\hyperlink{affil_4}{4},\orcidicon{0000-0002-0328-1640}}, 
C.~Brice\~{n}o\textsuperscript{\hyperlink{affil_13}{13},\orcidicon{0000-0001-7124-4094}}, 
C.~Burke\textsuperscript{\hyperlink{affil_2}{2},\orcidicon{0000-0002-7754-9486}}, 
D.~Bayliss\textsuperscript{\hyperlink{affil_4}{4},\hyperlink{affil_3}{3},\orcidicon{0000-0001-6023-1335}}, 
D.~Barrado\textsuperscript{\hyperlink{affil_9}{9},\orcidicon{0000-0002-5971-9242}}, 
E.M.~Bryant\textsuperscript{\hyperlink{affil_4}{4},\hyperlink{affil_3}{3},\orcidicon{0000-0001-7904-4441}}, 
D.J.A.~Brown\textsuperscript{\hyperlink{affil_3}{3},\hyperlink{affil_4}{4},\orcidicon{0000-0003-1098-2442}}, 
S.C.C.~Barros\textsuperscript{\hyperlink{affil_5}{5},\hyperlink{affil_11}{11},\orcidicon{0000-0003-2434-3625}}, 
F.~Bouchy\textsuperscript{\hyperlink{affil_10}{10},\orcidicon{0000-0002-7613-393X}}, 
D.A.~Caldwell\textsuperscript{\hyperlink{affil_14}{14},\orcidicon{0000-0003-1963-9616}}, 
D.M.~Conti\textsuperscript{\hyperlink{affil_15}{15},\orcidicon{0000-0003-2239-0567}}, 
R.F.~Díaz\textsuperscript{\hyperlink{affil_16}{16},\orcidicon{0000-0001-9289-5160}}, 
D.~Dragomir\textsuperscript{\hyperlink{affil_17}{17},\orcidicon{0000-0003-2313-467X}}, 
M.~Deleuil\textsuperscript{\hyperlink{affil_18}{18},\orcidicon{0000-0001-6036-0225}}, 
O.D.S.~Demangeon\textsuperscript{\hyperlink{affil_5}{5},\hyperlink{affil_11}{11},\orcidicon{0000-0001-7918-0355}}, 
C.~Dorn\textsuperscript{\hyperlink{affil_19}{19},\orcidicon{0000-0001-6110-4610}}, 
T.~Daylan\textsuperscript{\hyperlink{affil_2}{2},\hyperlink{affil_20}{20},\orcidicon{0000-0002-6939-9211}}, 
P.~Figueira\textsuperscript{\hyperlink{affil_21}{21},\hyperlink{affil_5}{5},\orcidicon{0000-0001-8504-283X}}, 
R.~Helled\textsuperscript{\hyperlink{affil_19}{19},\orcidicon{0000-0001-5555-2652}}, 
S.~Hoyer\textsuperscript{\hyperlink{affil_18}{18},\orcidicon{0000-0003-3477-2466}}, 
J.M.~Jenkins\textsuperscript{\hyperlink{affil_7}{7},\orcidicon{0000-0002-4715-9460}}, 
E.L.N.~Jensen\textsuperscript{\hyperlink{affil_22}{22},\orcidicon{0000-0002-4625-7333}}, 
D.W.~Latham\textsuperscript{\hyperlink{affil_6}{6},\orcidicon{0000-0001-9911-7388}}, 
N.~Law\textsuperscript{\hyperlink{affil_23}{23}}, 
D.R.~Louie\textsuperscript{\hyperlink{affil_24}{24},\orcidicon{0000-0002-2457-272X}}, 
A.W.~Mann\textsuperscript{\hyperlink{affil_23}{23},\orcidicon{0000-0003-3654-1602}}, 
A.~Osborn\textsuperscript{\hyperlink{affil_4}{4},\hyperlink{affil_3}{3},\orcidicon{0000-0002-5899-7750}}, 
D.L.~Pollacco\textsuperscript{\hyperlink{affil_4}{4},\orcidicon{0000-0001-9850-9697}}, 
D.R.~Rodriguez\textsuperscript{\hyperlink{affil_25}{25},\orcidicon{0000-0003-1286-5231}}, 
B.V.~Rackham\textsuperscript{\hyperlink{affil_2}{2},\hyperlink{affil_26}{26},\orcidicon{0000-0002-3627-1676}}, 
G.~Ricker\textsuperscript{\hyperlink{affil_2}{2},\orcidicon{0000-0003-2058-6662}}, 
N.J.~Scott\textsuperscript{\hyperlink{affil_7}{7},\orcidicon{0000-0003-1038-9702}}, 
S.G.~Sousa\textsuperscript{\hyperlink{affil_5}{5},\orcidicon{0000-0001-9047-2965}}, 
S.~Seager\textsuperscript{\hyperlink{affil_2}{2},\hyperlink{affil_26}{26},\orcidicon{0000-0002-6892-6948}}, 
K.G.~Stassun\textsuperscript{\hyperlink{affil_27}{27},\orcidicon{0000-0002-3481-9052}}, 
J.C.~Smith\textsuperscript{\hyperlink{affil_14}{14},\orcidicon{0000-0002-6148-7903}}, 
P.~Str\o{}m\textsuperscript{\hyperlink{affil_4}{4},\hyperlink{affil_3}{3},\orcidicon{0000-0002-7823-1090}}, 
S.~Udry\textsuperscript{\hyperlink{affil_10}{10},\orcidicon{0000-0001-7576-6236}}, 
J.~Villaseñor\textsuperscript{\hyperlink{affil_2}{2},\orcidicon{0000-0002-4625-8264}}, 
R.~Vanderspek\textsuperscript{\hyperlink{affil_2}{2},\orcidicon{0000-0001-6763-6562}}, 
R.~West\textsuperscript{\hyperlink{affil_3}{3},\hyperlink{affil_4}{4}}, 
P.J.~Wheatley\textsuperscript{\hyperlink{affil_3}{3},\hyperlink{affil_4}{4},\orcidicon{0000-0003-1452-2240}}, 
J.N.~Winn\textsuperscript{\hyperlink{affil_28}{28},\orcidicon{0000-0002-4265-047X}} 
}\\
\vspace{0.6cm}
\parbox{\textwidth}{
The authors' affiliations are shown in Appendix \ref{sec:affiliations}.}
}

\date{Accepted 07-Jan-2021}

\pubyear{2021}

\begin{document}
\label{firstpage}
\pagerange{\pageref{firstpage}--\pageref{lastpage}}
\maketitle

\begin{abstract}
We report the discovery of \Tplanet{} (\TTplanet{}), a transiting mini-Neptune exoplanet on a 2.5-day orbit around the solar-analogue \Tstar{} (\teff{}= $5730$K).
Using \tess{} photometry and \harps{} radial velocities gathered by the \textit{NCORES} program, we find \Tplanet{} has a radius of \Trpl{}\,\rearth{} and a mass of \TMpzero{}\,\mearth{}.
The resulting density of \Trhopgcmthree{}\,\gcm{} is significantly lower than would be expected from a pure-rock world; therefore, \Tplanet{} must be a mini-Neptune with a significant volatile atmosphere.
The high incident flux places it within the so-called radius valley; however, \Tplanet{} was able to hold onto a substantial (0.1-1\%) H-He atmosphere over its $\sim4$Gyr lifetime.
Through a novel simultaneous gaussian process fit to multiple activity indicators, we were also able to fit for the strong stellar rotation signal with period \Tperiod{}\,d from the RVs and confirm an additional non-transiting planet with a mass of \TMpone{}\,\mearth{} and a period of \TPone{}\,d.
\end{abstract}

\begin{keywords}
planets and satellites: detection -- stars: individual: HD110113
\end{keywords}



\section{Introduction}
Since its launch in 2018, NASA's \tess{} mission has attempted to detect small transiting planets around bright, nearby stars amenable to confirmation with radial velocity observations \citep{ricker2016transiting}.
The \harps{} spectrograph on the 3.6m telescope at La Silla, Chile \citep{2003Msngr.114...20M} has been deeply involved in this follow-up effort, beginning with its first detection, the hot super-Earth Pi Mensae c \citep{huang2018tess}, and continuing with the first multi-planet system \citep[TOI-125][]{quinn2019near,nielsen2020mass},

This unique combination of space-based photometry (which provides planetary radius) and precise radial velocities (which provide planetary mass) also allows for the determination of exoplanet densities, and, therefore, an insight into the internal structure of worlds outside our solar system.
These analyses have revealed a diversity of planet structures in the regime between Earth and Neptune, from high-density evaporated giant planet cores like TOI-849b \citep[5.2\,\gcm{}][]{armstrong2020remnant}, to low-density mini-Neptunes such as TOI-421\,c \citep{carleo2020multi}, as well as planets which follow a more linear track from rocky super-earths to Neptunes dominated by gaseous envelopes, such as the two inner planets orbiting $\nu^2$ Lupi \citep{kane2020transits} and TOI-735 \citep{cloutier2020pair,nowak2020carmenes}.

The detection of exoplanets with well-constrained physical parameters can also lead to the discovery of statistical trends within the planet population which encode information on planetary formation and evolution.
The "valley" seen around $1.8$\,\rearth{} in Kepler data \citep{fulton2017california, van2018asteroseismic} is one such feature.
According to current theory planets that first formed with gaseous envelopes within this valley have, due to heating from either their stars \citep[e.g. evaporation,][]{owen2017evaporation} or from internal sources \citep[e.g. core-powered mass loss,][]{ginzburg2018core}, lost those initial gaseous envelopes, thereby evolving to significantly smaller radii to become "evaporated cores".
By observing the physical parameters of small, hot exoplanets, the exact mechanisms of this process can be revealed.

In this paper, we present the detection, confirmation and RV characterisation of two exoplanets orbiting the star HD 110113 --- the hot mini-Neptune \Tplanet{} and the non-transiting \Tplanetc{}. 
The observations from which these planets were detected are described in section \ref{sect:observations}, while the analysis of that data is described in section \ref{sect:analysis}.
In section \ref{sect:discus} we discuss the validity of the outer planet RV signal (\ref{sect:planetc}), whether \Tstar{} is a solar analogue (\ref{sect:solaranal}) the internal structure and evaporation of planet b (\ref{sect:internal} \& \ref{sect:evap}), and potential future observations of the system (\ref{sect:future}).
We summarize our conclusions in section~\ref{sec:conclusion}.

\section{Observations}
\label{sect:observations}
\subsection{TESS photometry}
\Tstar{} was observed during \tess{} sector 10 with 2-minute cadence for 22.5 days, excluding a 2.5 day gap between \tess{} orbits to downlink data.
The lightcurve was extracted using the SPOC \citep[Science Processing Operations Centre;][]{jenkins2016tess} SAP (simple aperture photometry) pipeline.
It was then processed using the Pre-Search Data Conditioning \citep[PDC,][]{stumpe2012kepler,2012PASP..124.1000S,stumpe2014multiscale} pipeline, producing precise detrended photometry with typical precision of 150\,ppm/hr for this star, and then searched for exoplanetary candidates with the Transiting Planet Search \citep[TPS;][]{2010SPIE.7740E..0DJ}.
This identified a strong candidate with a period of 2.54\,d, a depth of only \Tdepth{} and a Signal to Noise Ratio (SNR) of 7.6.
Automated and human vetting subsequently designated this candidate a planet candidate and it was assigned \tess{} Object of Interest (TOI) 755.01. 

\begin{figure}
    \centering
    \includegraphics[width=\columnwidth, trim={0 0.8cm 1.8cm 0.1cm}]{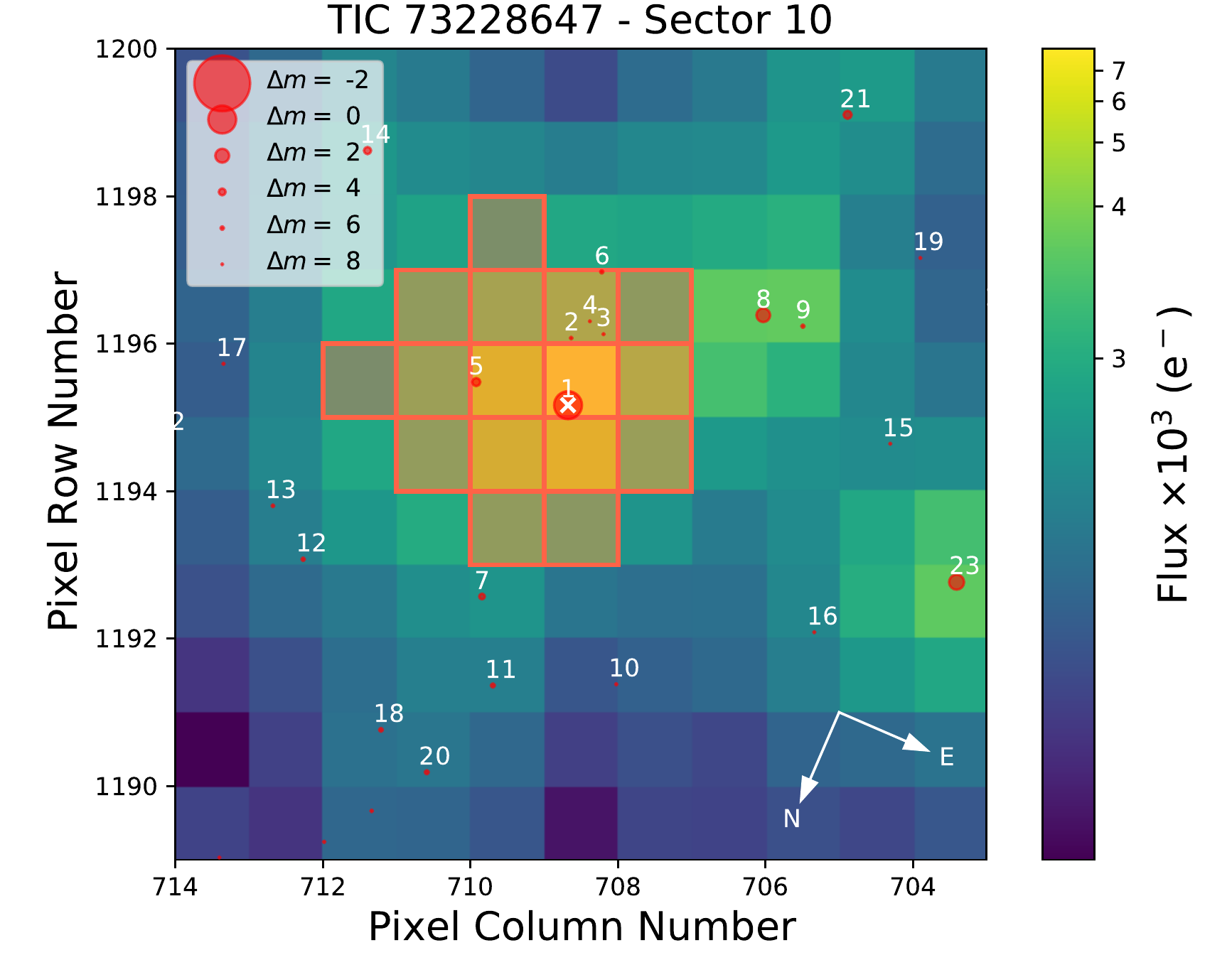}
    \caption{TESS photometric aperture plotted with \texttt{tpfplotter} \citep{2020A&A...635A.128A}. The default TESS aperture used by SAP is overplotted in red, and nearby stars down to $\Delta{\rm mag}=8$ from Gaia DR2 \citep{brown2018gaia} are plotted as red circles. The target \Tstar{} is marked with a white cross.}
    \label{fig:tpf}
\end{figure}

We inspected the \tess{} aperture using \texttt{tpfplotter} \citep[plotted in Figure \ref{fig:tpf};][]{2020A&A...635A.128A} to ensure no nearby contaminant stars could be causing the transit.
We found five stars within the aperture with contrast less than 8\,mag, with the brightest with a $\Delta{\rm mag}$ of only 3.5.
However, to cause the observed \Tdepth{} transit, this star would need to host eclipses of at least 1\%. 
Furthermore, being more than 1.2\,pix, and therefore almost one full-width-half-maximum (FWHM) of the point-spread function (PSF), away from the target star, we would expect to see a significant centroid shift. 
However, the SPOC data validation modelling \citep{Twicken:DVdiagnostics2018,Li:DVmodelFit2019} shows no such shift and suggests the transit occurs within 0.25 pixels from the target position\footnote{As shown by the SPOC DV report accessed at \url{https://mast.stsci.edu/api/v0.1/Download/file/?uri=mast:TESS/product/tess2019085221934-s0010-s0010-0000000073228647-00212_dvr.pdf}.}.
The other stars present are also $>1$ pixel away, and are increasingly fainter ($\Delta{\rm mag}$ of 6.9--7.9), requiring eclipse depths of 25--75\%.
Causing the observed transit with such a blend scenario therefore becomes increasingly unlikely given the flat-bottomed transit shape of TOI-755.01.
We conclude that a blend scenario from a known contaminant is unlikely, however we pursue additional photometry to confirm.

\subsection{Ground-based Photometric Follow-up}\label{sect:groundfu}
We observed a full transit of TOI-755.01 continuously for 443 minutes in Pan-STARSS $z$-short band on UTC 2020 March 13 from the LCOGT \citep{Brown:2013} 1-m network node at Cerro Tololo Inter-American Observatory.
The $4096\times4096$ LCOGT SINISTRO cameras have an image scale of $0\farcs389$ per pixel, resulting in a $26\arcmin\times26\arcmin$ field of view.
The images were calibrated by the standard LCOGT {\tt BANZAI} pipeline \citep{McCully:2018}.
Photometric data were extracted using {\tt AstroImageJ} \citep{Collins:2017}.
The mean stellar PSF in the image sequence had a FWHM of $2\farcs8$.
Circular apertures with radius $3\farcs1$ were used to extract the differential photometry.

The TOI-755 SPOC pipeline transit depth of 397\,ppm is too shallow to reliably detect with ground-based observations, so we instead checked for possible nearby eclipsing binaries (NEBs) that could be contaminating the irregularly shaped SPOC aperture that generally extends $\sim1\arcmin$ from the target star.
To account for possible contamination from the wings of neighboring star PSFs, we searched for NEBs out to $2\farcm5$ from the target star.
If fully blended in the SPOC aperture, a neighboring star that is fainter than the target star by 8.54 magnitudes in TESS-band could produce the SPOC-reported flux deficit at mid-transit (assuming a 100\% eclipse).
To account for possible delta-magnitude differences between TESS-band and Pan-STARSS $z$-short band, we searched an extra 0.5 magnitudes fainter (down to \textit{TESS}-band magnitude 18.5). 

The brightness and distance limits resulted in a search for NEBs in 90 Gaia DR2 stars, which includes all stars marked in red in Figure \ref{fig:tpf} and a further 67 contaminants with $\Delta{\rm mag}>8$.
We estimated the expected NEB depth in each neighboring star by taking into account both the difference in magnitude relative to TOI-755 and the distance to TOI-755 (to account for the estimated fraction of the star's flux that would be contaminating the TOI-755 SPOC aperture).
If the RMS of the 10-minute binned light curve of a neighboring star is more than a factor of 3 smaller than the expected NEB depth, we consider an NEB to be tentatively ruled out in the star over the observing window.
We then visually inspect each neighboring star's light curve to ensure no obvious eclipse-like signal.
The LCOGT data rule out possible contaminating NEBs at the SPOC pipeline nominal ephemeris and over a -$1.7\sigma$ to +$2.3\sigma$ ephemeris uncertainty window.
By process of elimination, we conclude that the transit is indeed occurring in TOI-755, or a star so close to TOI-755 that it was not detected by Gaia DR2, or the event occurred outside our observing window.

\subsection{Ground-based Archival Photometry}
Although detecting the transits of \Tplanet{} required precise space-based photometry, ground-based photometric surveys have observed \Tstar{} and can provide constraints on stellar variability, and therefore an independent measure of the stellar rotation period.

WASP-South was a wide-field array of 8 cameras forming the Southern station of the WASP transit-search survey \citep{2006PASP..118.1407P}. The field of \Tstar{} was observed over 150-night spans in each of 2007 and 2008, and then again over 2011 and 2012, acquiring a total of 30\,000 photometric data points. 
WASP-South was at that time equipped with 200-mm, f/1.8 lenses, observing with a 400--700 nm passband, and with a photometric extraction aperture of 48 arcsecs. 
There are other stars in the aperture around \Tstar{}, but the brightest has  $\Delta{\rm mag}=3$ while the others have $\Delta{\rm mag}>5$. 
Therefore any rotation signal is likely from \Tstar{}.
We searched the data for rotational modulations using the methods from \citet{2011PASP..123..547M}.

\begin{figure}
\centering
\includegraphics[width=\columnwidth]{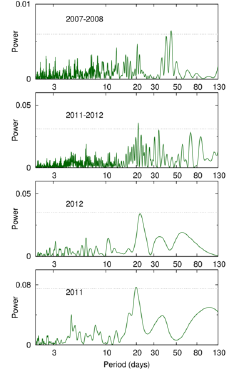}
  \caption{Periodograms of the WASP-South data for TOI-755. The top panel shows data from 2007 \&\ 2008 combined, with a significant 42-d periodicity. The lower panels show data from 2011 \&\ 2012, separately and combined, which show more strongly a periodicity of 21-d. The dotted horizontal lines are the estimated 1\%-likelihood false-alarm levels.}
\label{fig:wasp}
\end{figure}

The data from 2011 and 2012 show a modulation at a period of 21\,$\pm$\,2 d (see Figure \ref{fig:wasp}).
This is significant at the 1\%\ false-alarm level with an amplitude of 2 mmag, both in each year separately, and when the data from the two years are combined.
The data from 2007 and 2008 combined show a significant modulation at twice this period of $\sim42$\,$\pm$\,4 d.
There is also power near 21 d in the 2007/2008 periodogram, but it is not significant in its own right.
Although it would seem more likely from these data alone that the rotational period of \Tstar{} is 42\,$\pm$\,4 days, with the $21$\,d period coming from the first harmonic, the WASP data cannot on its own distinguish between these two possible periods.

\subsection{High-resolution imaging}
\begin{figure}
    \centering
    \includegraphics[width=\columnwidth, trim={0 2cm 0.6cm 1.2cm}]{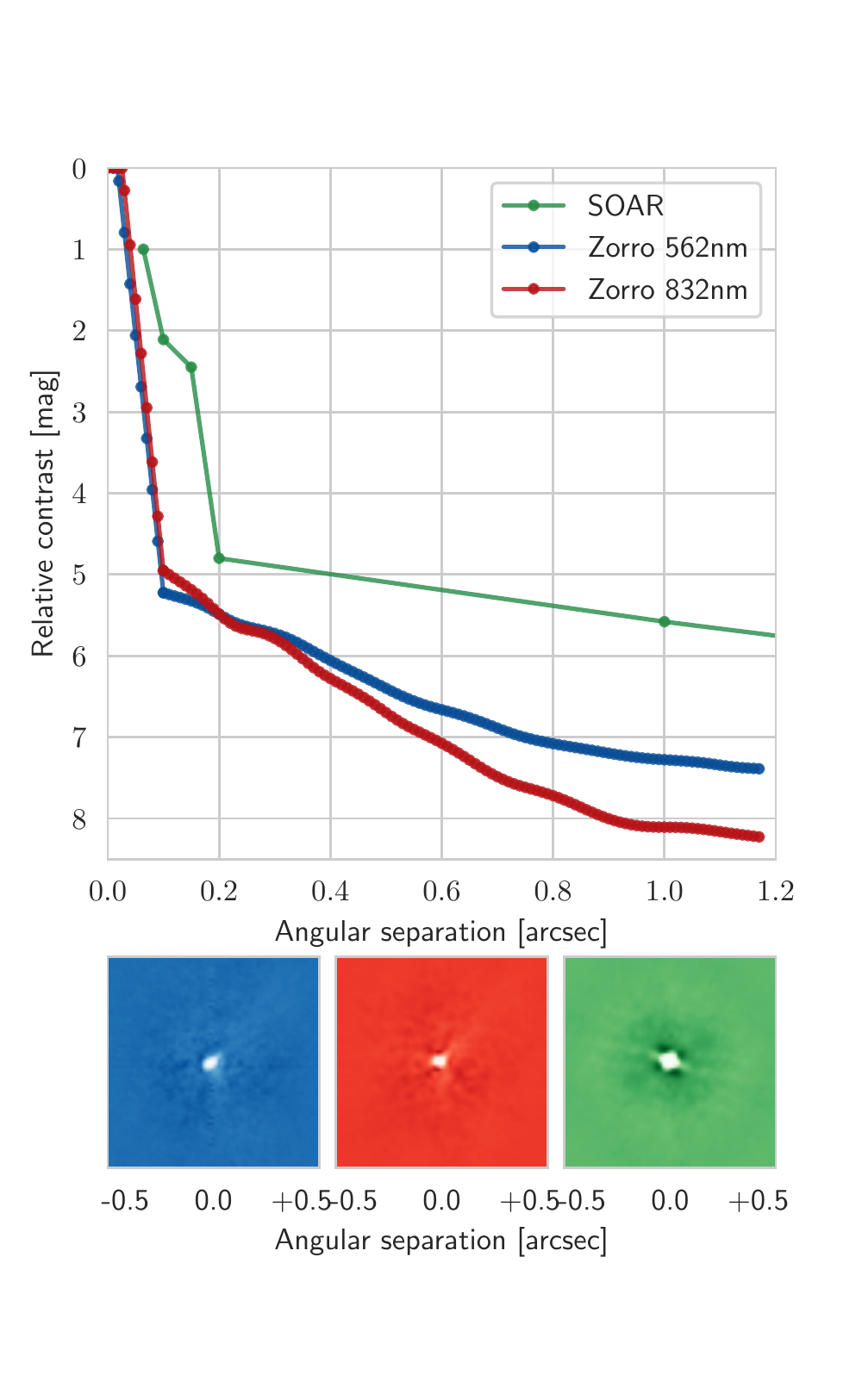}
    \caption{Contrast curves and images from Gemini/Zorro (blue \& red for 562 and 832nm respectively), and SOAR (green).}
    \label{fig:imaging}
\end{figure}

High-angular-resolution imaging is needed to search for nearby sources that can contaminate the \tess{} photometry, resulting in an underestimated planetary radius, or that can be the source of astrophysical false positives, such as background eclipsing binaries. 
Through the TESS Follow-Up Program (TFOP), three such images were obtained across two telescopes, with the results shown in Figure \ref{fig:imaging}.

\subsubsection{SOAR}
We searched for stellar companions to TOI-755 with speckle imaging on the 4.1-m Southern Astrophysical Research (SOAR) telescope \citep{tokovinin2018ten} on 14 July 2019 UT, observing in Cousins I-band, a similar visible bandpass as \tess{}. More details of the observation are available in \citet{ziegler2020vizier}. The $5\sigma$ detection sensitivity and speckle auto-correlation functions from the observations are shown in Figure \ref{fig:imaging}. No nearby stars with magnitudes brighter than $I=16$ were detected within 3\arcsec{} of \Tstar{} in the SOAR observations.

\subsubsection{Gemini/Zorro}
High-resolution speckle interferometric images of \Tstar{} were obtained on 14 January 2020 UT using the Zorro instrument mounted on the 8-meter Gemini South telescope located on the summit of Cerro Pachon in Chile. 
Zorro simultaneously observes in two bands, (832\,nm \& 562\,nm with widths of 40 \& 54\,nm respectively), obtaining diffraction-limited images with inner working angles 0.017 and 0.028 arcsec, respectively.
The observation consisted of 3-minute sets of 1000$\times$0.06-second images. All the images were combined and subjected to Fourier analysis, leading to the production of final data products, including speckle reconstructed imagery \citep[see][]{2011AJ....142...19H}.
Figure \ref{fig:imaging} shows the $5\sigma$ contrast curves in both filters for the Zorro observation and includes an inset showing the 832 nm reconstructed image.
The resulting contrast limits reveal that \Tstar{} is a single star to contrast limits of 5 to 8 magnitudes, ruling out most main sequence companions to the star within the spatial limits of $\sim$11 to 320\,au (for $d=106.3$pc).

\subsection{HARPS High Resolution Spectroscopy}
Over the course of two observing seasons in 2018 and 2019, a total of 114 high-resolution spectra were taken with the High Accuracy Radial velocity Planet Searcher \citep[HARPS,][]{Pepe:2002,2003Msngr.114...20M} on the ESO 3.4m telescope at La Silla, Chile.
These spectra were taken as part of the \textit{NCORES} program (PI:Armstrong, 1102.C-0249) designed to specifically study the internal structure of hot worlds.

We used the high-accuracy mode of HARPS with a $1\arcsec$ science fibre on the star and a second on-sky fibre monitoring the background flux during exposure. 
The nominal exposure time was 1800 seconds, with a few exceptions of slightly longer or shorter integration, depending on observing conditions and schedule.

Spectra and RV information were extracted using the 
offline HARPS data reduction pipeline hosted at Geneva Observatory. 
We use a flux template matching a G1 star to correct the continuum-slope in each echelle order. 
The spectra were cross correlated with a binary G2 mask to derive the cross correlation function (CCF) \citep{1996A&AS..119..373B}, on which we fit a Gaussian function to obtain RVs, FWHM and contrast. 
Additionally, we compute the bisector-span \citep{2001A&A...379..279Q} of the CCF and spectral indices tracing chromospheric activity \citep{2011A&A...534A..30G,2009A&A...495..959B}.

We reach a typical SNR per pixel of 75 (order 60, 631nm) in individual spectra, corresponding to an RV error of 1.41\ms{}.
The HARPS spectra and derived RVs were accessed and downloaded through the DACE portal hosted at the University of Geneva \citep{2015ASPC..495....7B} under the target name HD 110113\footnote{\url{https://dace.unige.ch/radialVelocities/?pattern=HD110113}}.

\section{Analysis}\label{sect:analysis}
\subsection{Stellar Parameters}

\begin{table}
    \centering
    \begin{tabular}{lc|lc}
        \hline
        \hline
        Parameter & Value & Parameter & Value \\
        \hline
        \hline
        TOI ID & \TTstar & R.A. [$^{\circ}$] & $190.0365636$\,$^{a}$ \\
        TIC ID & 73228647 $^{b}$ & R.A. [hms] & 12:40:08.78 $^{a}$ \\
        HD & \Tstar & Dec. [$^{\circ}$] & $-44.3120777$\,$^{a}$\\
        HIP & HIP 61820 & Dec. [dms] & -44:18:43.48 $^{a}$ \\
        Gaia ID & {\scriptsize \TGAIAid $^{a}$} & $\delta{\rm RA}$ [mas\,${\rm yr}^{-1}$] & $-3.72 \pm 0.1$\,$^{a}$ \\
        $\pi{} [{\rm mas}]$ & $9.38 \pm 0.036$\,$^{a}$ & $\delta{\rm DEC}$ [mas\,${\rm yr}^{-1}$] & $-13.68 \pm 0.12$\,$^{a}$ \\
        $d$ [pc] & $ 106.3\pm0.72$\,$^{e}$ & $R_s$ [\rsun{}] & \Trstar{}$^{e}$ \\
        B & $10.71\pm0.032$\,$^{c}$ & $M_s$ [\msun{}] & \Tmstar{}$^{e}$ \\
        V & $10.063\pm0.027$\,$^{c}$ &  \logg{} & \Tlogg{}$^{e}$ \\
        Gaia $G$ & $9.91\pm0.0004$\,$^{a}$ & \teff{} [K] & \TTeff{}$^{e}$ \\
        TESS mag & $9.4628\pm0.006$\,$^{b}$ & \feh{} & \TFeH{}$^{e}$ \\
        J & $8.903 \pm 0.037$\,$^{c}$ & \vsini{} [\kms{}]& $1.74\pm0.15$\,$^{e}$\\
        H & $8.594 \pm 0.063$\,$^{c}$ & $P_{\rm rot}$ [d] & \Tperiod{}$^{f}$\\
        K & $8.502 \pm0.024$\,$^{c}$ & Age $[{\rm Gyr}]$ & \Tstarage{} $^{g}$ \\
        \hline
        \hline
    \end{tabular}
    \caption{Stellar parameters.
    $^{a}$\,From Gaia DR2\citep{brown2018gaia}. $^{b}$\,From the \tess{} Input Catalogue v8 \citep{stassun2019revised}. $^{c}$\,Johnson magnitudes from APASS \citep{apass}. $^{d}$\,From 2MASS \citep{skrutskie2006two}. $^{e}$\,Derived from \harps{} spectra and archival data - see sect \ref{sect:starpars}. $^{f}$\,Determined using the GP fit to activity indicators and RVs as described in \ref{sect:rvs}. $^{g}$\,Derived from $[\rm{Y}/\rm{Al}]$ abundance-age relation as described in sect \ref{sect:abunds}.}
    \label{tab:starpars}
\end{table}

\subsubsection{Global Stellar Parameters}\label{sect:starpars}
The star's effective temperature (\teff{}), surface gravity (\logg{}), and metallicity (\feh{}) were derived using a recent version of the MOOG code \citep{1973ApJ...184..839S} and a set of plane-parallel ATLAS9 model atmospheres  \citep{Kurucz-93}. The analysis was done in LTE. 
The methodology used is described in detail in \citet{2011A&A...533A.141S} and \citet{2013A&A...556A.150S}. 
The full spectroscopic analysis is based on the Equivalent Widths (EWs) of 233 Fe \rom{1} and 34 Fe \rom{2} weak lines by imposing ionization and excitation equilibrium. 
The line-list used was taken from \citet{2008A&A...487..373S}.
We obtained resulting parameters of \teff{}=$5732\pm16$K, \logg{}$=4.46\pm0.05$ and \feh{}$=0.14\pm0.02$.
To account for potential systematic uncertainties, we increased the error bars to $50$K and 0.05 dex for \teff{} and \logg{} respectively.

To constrain the physical stellar parameters of \Tstar{} given the observed information, we applied three techniques.

The first technique was to use the main-sequence calibrations of \citet{2010A&ARv..18...67T} which derive $R_s$ and $M_s$ using polynomial functions of \teff{}, \logg{} and \feh{}, which are built using the observed properties of calibration stars. 
Uncertainties were propagated using 10000 Monte Carlo draws and the mass was corrected using the calibration of \citet{Santos-13}. 
This produced a mass and radius of \Tmstartorres{}\,\msun{} and \Trstartorres{}\,\rsun{} respectively, although \citet{2010A&ARv..18...67T} suggest minimum uncertainties of 0.06\msun{} and 0.03\rsun{} respectively.

The second was using theoretical isochrones \citep[MIST,][]{2016ApJ...823..102C} as well as observed properties (e.g. colours) to constrain stellar parameters, which we performed using \texttt{isoclassify} \citep{2017zndo....573372H,2020AJ....159..280B}.
Inputs included the derived spectral properties \teff{}, \logg{} and \feh{}, as well as archival data for \Tstar{} including {\it APASS} B \& V magnitudes \citep{apass}, {\it Gaia} parallax, Gp, Rp, Bp and luminosity \citep{brown2018gaia}, \textit{SkyMapper} ugriz observations \citep{2020arXiv200810359O} and {\it 2MASS} JHK observations \citep{skrutskie2006two}.
This resulted in a mass \& radius of \Tmassiso{}\,\msun{} and \Tradiso{}\,\rsun{} respectively.
The well-constrained nature of the input measurements mean that we are limited by the gridsize of the theoretical isochrones, which despite an initial array of more than 3 million points, resulted in only 112 samples within all available constraints.

\begin{figure}
    \centering
    \includegraphics[width=\linewidth,trim=100 75 95 95,clip]{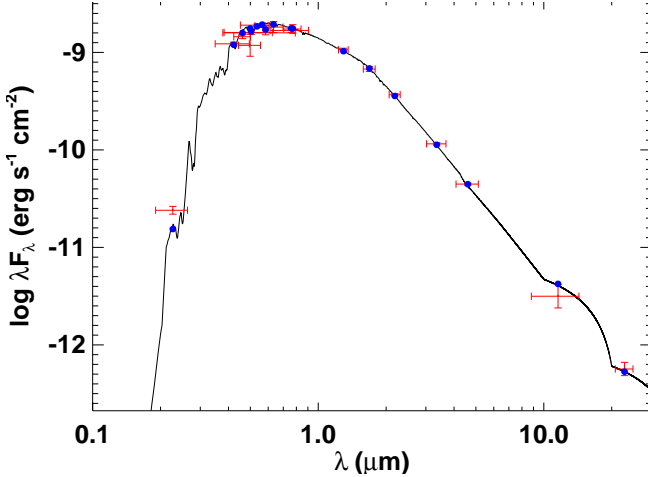}
    \caption{Spectral energy distribution of \Tstar. Red symbols represent the observed photometric measurements, where the horizontal bars represent the effective width of the passband. Blue symbols are the model fluxes from the best-fit Kurucz atmosphere model (black).}
    \label{fig:sed}
\end{figure}

As a final independent determination of the basic stellar parameters for \Tstar, we performed an analysis of the broadband spectral energy distribution (SED) of the star together with the {\it Gaia\/} DR2 parallax \citep[adjusted by $+0.08$~mas to account for the systematic offset reported by][]{StassunTorres:2018}, in order to determine an empirical measurement of the stellar radius, following the procedures described in \citet{Stassun:2016,Stassun:2017,Stassun:2018}. We pulled the $B_T V_T$ magnitudes from {\it Tycho-2}, the $BVgri$ magnitudes from {\it APASS}, the $JHK_S$ magnitudes from {\it 2MASS}, the W1--W4 magnitudes from {\it WISE}, the $G G_{\rm BP} G_{\rm RP}$ magnitudes from {\it Gaia}, and the NUV magnitude from {\it GALEX}. Together, the available photometry spans the full stellar SED over the wavelength range 0.2--22~$\mu$m (see Figure~\ref{fig:sed}).  

We performed a fit using Kurucz stellar atmosphere models, with the $T_{\rm eff}$, [Fe/H] and $\log g$ adopted from the spectroscopic analysis. The only additional free parameter is the extinction ($A_V$), which we restricted to the maximum line-of-sight value from the dust maps of \citet{Schlegel:1998}. The resulting fit is very good (Figure~\ref{fig:sed}) with a reduced $\chi^2$ of 1.4 and best-fit $A_V = 0.03 \pm 0.03$. Integrating the (unreddened) model SED gives the bolometric flux at Earth, $F_{\rm bol} = 2.597 \pm 0.091 \times 10^{-9}$ erg~s$^{-1}$~cm$^{-2}$. Taking the $F_{\rm bol}$ and $T_{\rm eff}$ together with the {\it Gaia\/} DR2 parallax gives the stellar radius, $R_\star = 0.968 \pm 0.018 R_\odot$. In addition, we can use the $R_\star$ together with the spectroscopic $\log g$ to obtain an empirical mass estimate of $M_\star = 0.99 \pm 0.08 M_\odot$.

Taken together, all the stellar parameters as derived above are consistent, and all suggest that \Tstar{} is a solar analogue with mass and radius very close to the Sun.
As the SED radius measurement is least affected by sample size or systematic uncertainty, we assume this as a final radius.
Similarly, the mass obtained from the \logg{} and the SED-derived \rstar{} ($0.99 \pm 0.08 M_\odot$) is nearly identical to that from the MR relationship (\Tmstartorres{}\,\msun{}), suggesting they converge on the same value.
We therefore use the mass as defined from the offset-corrected \citet{2010A&ARv..18...67T} calibrations, with the uncertainty inflated to reflect the typical systematic error ($0.06M_{\odot}$).

To compute the \vsini{} from the FWHM, we used the relations of \citet{dos2016solar}, who studied the \harps{} spectra of a large number of solar twins.
We used this to first estimate the $v_{\rm macro}$ from the \teff{} and \logg{} derived in section \ref{sect:starpars} ($3.64\pm0.1$\,\kms{}), and then combined this with the measured FWHM to estimate a \vsini{} of $1.74\pm0.15$\,\kms{}, although the uncertainties here may be underestimated due to systematic uncertainties.
Using the calculated $R_s$, this corresponds to a maximum rotation period ($P_{\rm max}$) of $28\pm3$\,d, assuming an aligned system.

\subsubsection{Chemical abundances} \label{sect:abunds}

Stellar abundances of the elements were also derived using the same tools and models as for stellar parameter determination as well as using the classical curve-of-growth analysis method assuming local thermodynamic equilibrium. Although the EWs of the spectral lines were automatically measured with ARES, for the elements with only two to three lines available we performed careful visual inspection of the EWs measurements. For the derivation of chemical abundances of refractory elements, we closely followed the methods described in the literature \citep[e.g.][]{Adibekyan-12, Adibekyan-15, Delgado-14, Delgado-17}. Abundances of the volatile elements, O and C, were derived following the method of \cite{Delgado-10, 2015A&A...576A..89B}. Since the two spectral lines of oxygen are usually weak and the 6300.3\,\AA{} line is blended with Ni and CN lines, the EWs of these lines were manually measured with the task \texttt{splot} in IRAF. Lithium and sulfur abundances were derived by performing spectral synthesis with MOOG, following the works by \citet{Delgado-14} and \citet{Costa_Silva2020} respectively. Both abundance indicators are very similar to the solar values.
All the [X/H] ratios are obtained by doing a differential analysis with respect to a high S/N solar (Vesta) spectrum from HARPS. The stellar parameters and abundances of the elements are presented in Table \ref{tab:abunds}. 

We find that the [X/Fe] ratios of most elements are close to solar as expected for a star with this metallicity whereas [O/Fe] and [C/Fe] are slightly subsolar, since these ratios tend to slightly decrease above solar metallicity \cite[e.g.][]{Bertrandelis-15,Franchini2020}. Moreover, we used the chemical abundances of some elements to derive ages through the so-called chemical clocks (i.e. certain chemical abundance ratios which have a strong correlation with age). We applied the 3D formulas described in \citet{Delgado-19}, which also consider the variation in age produced by the effective temperature and iron abundance. The chemical clocks [Y/Mg], [Y/Zn], [Y/Ti], [Y/Si], and [Y/Al] were derived. We selected the [Y/Al] age, \Tstarage{}, as the representative age, as it is consistent with all others and has the smallest uncertainty.

\begin{table}
    \centering
    \begin{tabular}{lcc}
        \hline
        \hline
        Parameter & Value & Error \\
        \hline
        \hline
        \multicolumn{3}{c}{\it Abundances}\\
        A(Li) & $1.09$ & $0.08$ \\
        $[\rm{Fe}/\rm{H}]$ & $0.14$ & $0.02$ \\
        $[\rm{S}/\rm{H}]$ & $0.03$ & $0.04$ \\
        $[\rm{Na}/\rm{H}]$ & $0.141$ & $0.038$ \\
        $[\rm{Mg}/\rm{H}]$ & $0.129$ & $0.021$ \\
        $[\rm{Al}/\rm{H}]$ & $0.105$ & $0.014$ \\
        $[\rm{Si}/\rm{H}]$ & $0.097$ & $0.022$ \\
        $[\rm{Ca}/\rm{H}]$ & $0.092$ & $0.062$ \\
        $[\rm{Ti}/\rm{H}]$ & $0.140$ & $0.030$ \\
        $[\rm{Cr}/\rm{H}]$ & $0.156$ & $0.032$ \\
        $[\rm{Ni}/\rm{H}]$ & $0.130$ & $0.024$ \\
        $[\rm{O}/\rm{H}]$ & $-0.012$ & $0.083$ \\
        $[\rm{C}/\rm{H}]$ & $0.032$ & $0.012$ \\
        $[\rm{Cu}/\rm{H}]$ & $   0.116 $ & $  0.016 $ \\
        $[\rm{Zn}/\rm{H}]$ & $   0.050 $ & $  0.012 $ \\
        $[\rm{Sr}/\rm{H}]$ & $   0.170 $ & $  0.073 $ \\
        $[\rm{Y}/\rm{H}]$ & $    0.170 $ & $  0.039 $ \\
        $[\rm{Zr}/\rm{H}]$ & $   0.152 $ & $  0.045 $ \\
        $[\rm{Ba}/\rm{H}]$ & $   0.123 $ & $  0.047 $ \\
        $[\rm{Ce}/\rm{H}]$ & $   0.120 $ & $  0.051 $ \\
        $[\rm{Nd}/\rm{H}]$ & $   0.135 $ & $  0.056 $ \\
        \hline
        \multicolumn{3}{c}{\it Derived Abundance Ratios}\\
        Mg/Si & $   1.32 $ & $ 0.09 $ \\
        Fe/Si & $   1.08 $ & $ 0.07 $ \\
        Mg/Fe & $   1.23 $ & $ 0.08 $ \\
        \hline
        \multicolumn{3}{c}{\it Ages}\\
        $[\rm{Y}/\rm{Mg}]$ Age [Gyr]  & $ 4.09 $ & $ 0.75 $ \\
        $[\rm{Y}/\rm{Ti}]$ Age [Gyr]  & $ 4.09 $ & $ 0.95 $ \\
        $[\rm{Y}/\rm{Zn}]$ Age [Gyr]  & $ 3.29 $ & $ 0.77 $ \\
        $[\rm{Y}/\rm{Si}]$ Age [Gyr]  & $ 3.95 $ & $ 0.86 $ \\
        $[\rm{Y}/\rm{Al}]$ Age [Gyr]  & $ 4.00 $ & $ 0.54 $ \\
        \hline
        \hline
    \end{tabular}
    \caption{Derived stellar abundances. [Y/X] based ages using the 3D formula of \citet{Delgado-19} (Table 10: age \&  $a + b \times$\teff{}$+ c \times$\feh{}$ + d \times$\ymg{}).}
    \label{tab:abunds}
\end{table}

\subsection{Combined modelling of RV \& Photometry}

\subsubsection{Treatment of Radial Velocities}\label{sect:rvs}
\begin{figure}
	\includegraphics[width=\columnwidth, trim={0.3cm 1.1cm 0.8cm 1.3cm}]{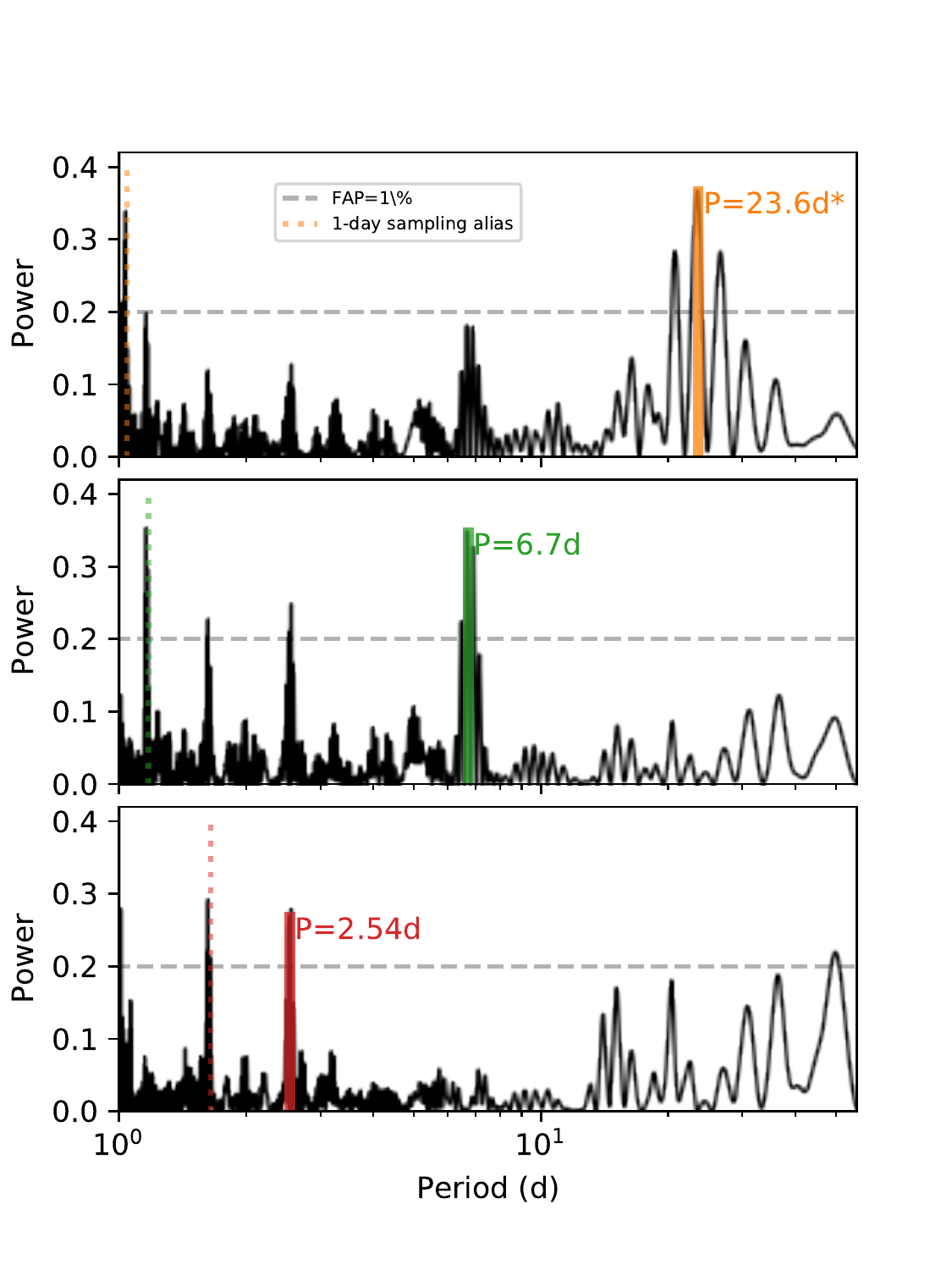}
    \caption{Periodograms of RVs after linear decorrelation with S-index and FWHM. The upper panel shows the raw periodogram, while subsequent panels show the periodogram after the removal of the previously marked peak. The 2.54\,d peak is accompanied by a significant peak at the 1-day sampling alias (1.65\,d), but the knowledge of a 2.54\,d planet in the \tess{} photometry breaks this degeneracy. The remaining peaks in the final periodogram are likely due to sampling aliases associated with the $\sim60$\,d span of observations.}
    \label{fig:rv_decorr}
\end{figure}

All activity indicators showed clear signs of stellar variability, likely due to the presence of starspots.
To remove this stellar activity, we first turned to linear decorrelation of the RV signal using activity indicators.
The FWHM and S-index showed the clearest rotational signals, so we selected these and used the decorrelation technique provided with the DACE spectroscopy Python package \citep{2015ASPC..495....7B}\footnote{\url{https://dace.unige.ch/tutorials/?tutorialId=34}}.
Despite this decorrelation removing much of the stellar variability signal, the peak at $\sim22$d remained the single strongest signal in the radial velocity time series (see Figure \ref{fig:rv_decorr}).
To remove the rotation signal at $23.68\pm0.08$\,d, we fitted a 5-parameter Keplerian model (with eccentricity $e$, argument of periastron $\Omega$, \& semi-amplitude $K$ as free parameters, with period $P$ and time of transit $t_0$ constrained from the periodogram).
The next strongest signals were at $6.73\pm0.03$\,d and $2.541\pm0.0008$\,d with amplitudes of $3.88\pm0.31$\,\ms{} and $2.55\pm0.31$\,\ms{} respectively. This was followed by signals on longer periods, which are most likely spurious due to rotational and observational aliases.

Although this linear decorrelation and Keplerian-fitted rotation period was able to reveal the planetary RV signals, stellar variability cannot in general be modelled as a Keplerian.
Instead we turned to a Gaussian process (GP) to model the impact of rotation on the RVs.
GPs have frequently been used in the analysis of radial velocities affected by activity \citep[e.g.][]{2014MNRAS.443.2517H,2019A&A...627A..43D}.
One GP kernel well-suited to stellar rotation is a mix of simple harmonic oscillator (SHO) terms corresponding to $P_{\rm rot}$ and $P_{\rm rot}/2$, which we built using \texttt{exoplanet} and \texttt{celerite} packages\footnote{We used the \texttt{exoplanet.gp.terms.RotationTerm} implementation}.

In order to limit the impact of the GP on the planetary RV signal, we fitted activity indicators and RV time-series simultaneously with the same GP kernel, as these should follow the same underlying variations with the exception of planetary reflex motion.
A similar approach was previously used by \citet{grunblatt2015determining} to model stellar variability in the Kepler-78b system, and by \citet{2020A&A...639A..77S} to find an outer candidate orbiting Proxima Centauri.
By explicitly linking the variation found across activity indicators and RVs, this method has the same effect as "training" a GP on an activity indicator \citep[e.g.][]{2019A&A...627A..43D}.
However, it avoids having to run multiple models consecutively and transfer the output PDF of a training sample into a second model---a process which loses information intrinsic to the likely non-Gaussian distributions of the GP hyper-parameters as well as information about the correlations between parameters.
This technique also enables the use of multiple time-series.
In this case, we chose S-index and FWHM to co-fit the covariance function with the RVs, as these showed the clearest rotation signal.

To achieve this, the hyper-parameters for rotation period, mix factor between $P_{\rm rot}$ and $P_{\rm rot}/2$ terms, signal quality ($Q$), and the difference in signal quality between modes ($\Delta Q$) were kept constant between S-index, FWHM and RV time-series, while the signal amplitude and mean, which are not shared across parameters, were set as separate parameters.
For each time-series we also used a jitter term to model noise not included by measurement errors and to prevent GP over-fitting.
All hyper-parameters were given broad priors, although the rotation period was constrained to the value obtained from a Lomb-Scargle periodogram \citep{Lomb1976, Scargle1982} with a standard deviation of 20\%.
All parameter priors are listed in Table \ref{tab:planetparlong}.

We also noted that the FWHM errors produced by the \harps{} pipeline appeared over-estimated---more than twice the estimated error derived from the median absolute difference between measurements.
Therefore the FWHM errors were multiplied by a factor of 0.4386 such that the median error matched the point-to-point RMS as calculated from the median absolute difference.

While we used the GPs to model the covariance between points in each timeseries, a mean function is also required to calibrate the average value over time, which we applied separately to each of the three timeseries.
A 2-parameter (i.e. linear) trend term was included to model potential long-term drift in the RVs, although the resulting gradient was not significant ($-0.14\pm0.73$\,\ms{}d$^{-1}$).
Single-parameter mean values were included to model the offset of S-index and FWHM from zero.

\begin{figure*}
	\includegraphics[width=\textwidth, trim={0.85cm 0.8 1.9cm 0.4cm}]{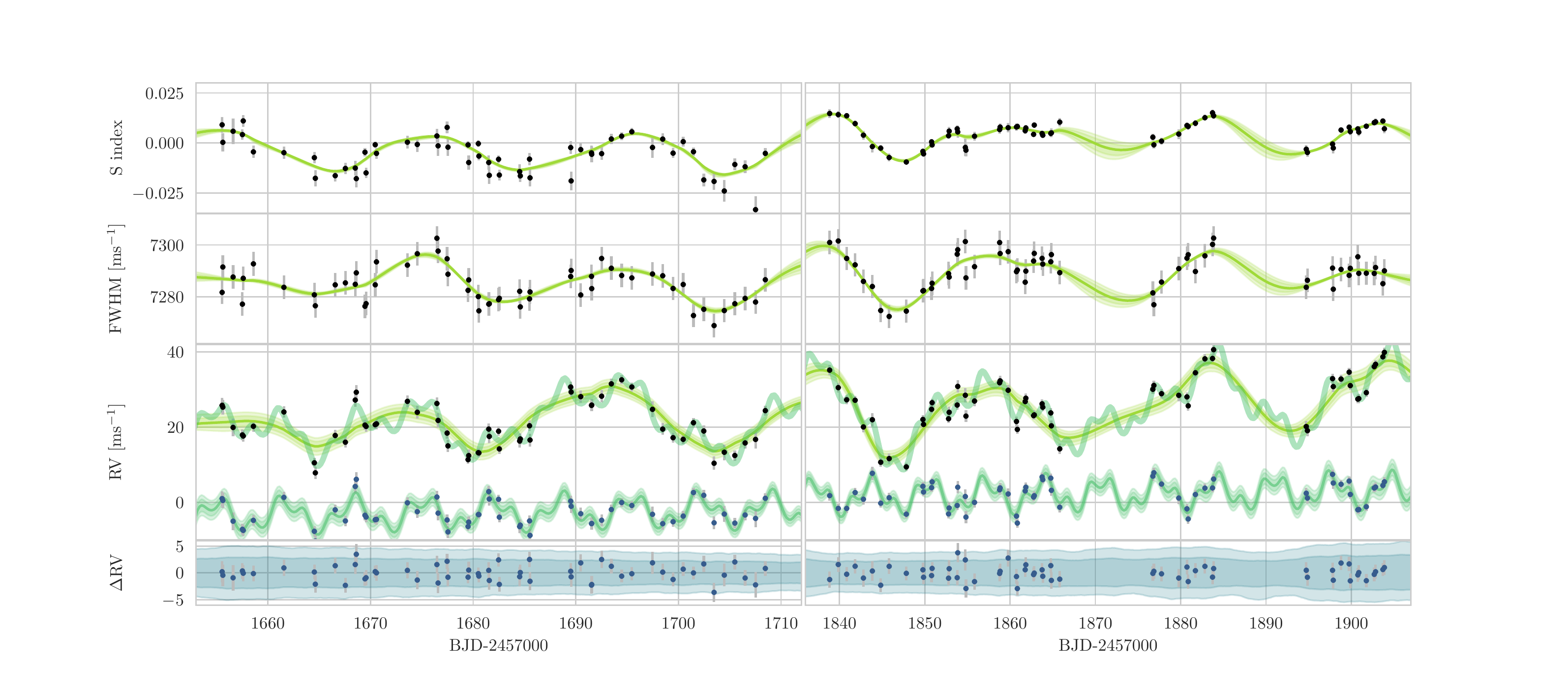}
    \caption{S-index, FWHM and RV timeseries of HD110113 for two seasons of HARPS monitoring, with GP models and 2-sigma uncertainty regions overplotted in green. Below the raw RV timeseries is the GP-removed RV timeseries, with the modelled planetary reflex motion and background trend (turquoise). At the very bottom the full model residuals are shown, with an RMS of only 1.31\,\ms{}---extremely close to the median HARPS measurement uncertainty (1.36\,\ms{}).}
    \label{fig:RVs}
\end{figure*}

\begin{figure}
	\includegraphics[width=\columnwidth, trim={0.1cm 0.8cm 1.0cm 0.85cm}]{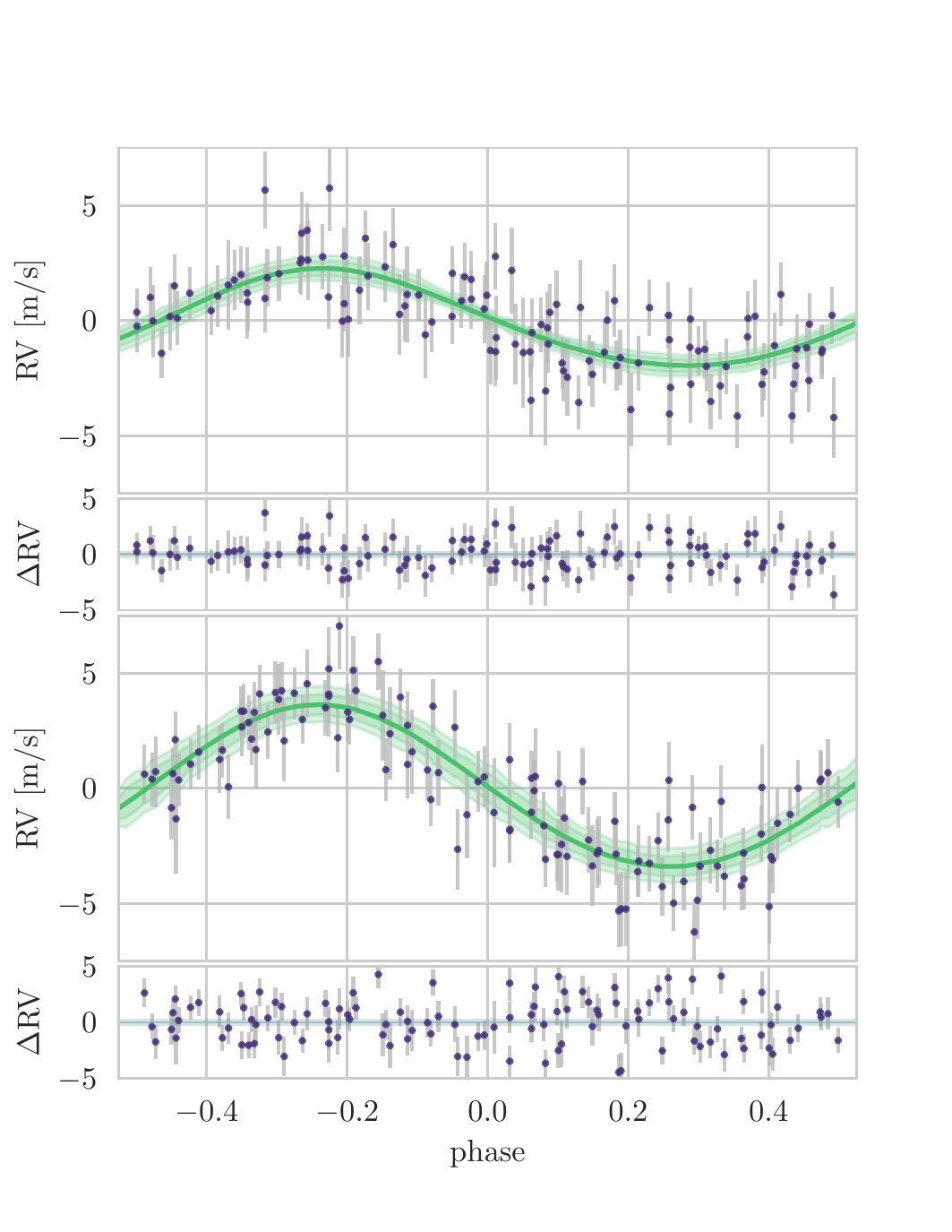}
    \caption{Phase-folded RVs (with the best-fit GP model, linear trend, and the other planetary signal removed) for \Tplanet{} (top) and \Tplanetc{} (bottom), with model-subtracted residuals. }
    \label{fig:phase_fold_rvs}
\end{figure}

\subsubsection{Treatment of Photometry}
We downloaded the \texttt{PDC\_SAP} lightcurve from the Mikulski Archive for Space Telescopes (MAST).
As high-resolution imaging revealed no close stellar neighbours missed by e.g. the \tess{} input catalogue \citep{stassun2019revised}, we made the assumption that the PDC-extracted and dilution-corrected lightcurve for this target was accurate.

We then normalised the \texttt{PDC\_SAP} timeseries by its median and masked anomalous flux points from the timeseries by cutting data more than $4.2\sigma$ different from both preceding and succeeding neighbours.

We initially tried to use the same \texttt{celerite} GP kernel to predict both RV and photometric time-series deviations.
This proved to not be possible, likely because the effect of stellar variability on photometry is not necessarily at the same timescale as for RVs \citep{10.1111/j.1365-2966.2011.19960.x}.
Similarly, although a Lomb-Scargle periodogram of the raw \tess{} lightcurve does show a peak with a period around 25d, the processed \texttt{PDC\_SAP} lightcurve is flat, likely as variability on the order of a \tess{} orbit ($\sim 14$\,d) is removed during processing.

The remaining variability is therefore likely to be the result of stellar granulation, which is well-suited to be modelled with a single GP SHO kernel with quality $Q=1/\sqrt{2}$ \citep{2020A&A...634A..75B, exoplanet:foremanmackey17}.
To produce the initial hyperparameters ($\omega_0$ \& $S_0$) and priors for the combined analysis and reduce the possibility of the GPs attempting to model the transits themselves, we first fitted this GP to the photometry with planetary transits cut. 
The interpolated posterior distributions from this analysis then provided the priors for the combined analysis.
A jitter term was also included to model the effect of high-frequency noise not fully encapsulated by the photon noise (e.g. stellar \& spacecraft jitter).

We modelled the limb darkening using two approaches: one where limb darkening is a free parameter, reparameterised using the approach of \citet{kipping2013efficient} and fitted to the transit with uninformative priors that cover the physical parameter space; and another where the expected theoretical limb darkening parameters for the star as generated by \citet{claret2017limb} are used as priors for the analysis.
We found the resulting distributions to be consistent, and chose to use the second, constrained approach in the final modelling. 
This used a normal prior with the mean, $\mu$, set from the theoretical parameter and $\sigma$ set as 0.1 which we chose instead of the uncertainty found when propagating the stellar parameters through the \citet{claret2017limb} relation, which was likely too contsraining and did not account for systematic uncertainties.
The radius ratio $R_p/R_s$ was treated using the log amplitude to avoid negative values, and b was reparameterised with $R_p/R_s$ following the \texttt{exoplanet} implementation of \citet{espinoza2018efficient}.

As ground-based photometry was not precise enough to observe a transit (see Sect. \ref{sect:groundfu}), we restrict this analysis to only the \tess{} photometry and \harps{} spectroscopy.

\subsubsection{Combined Model}
We modelled full Keplerian orbits for the two planets, with eccentricity priors according to the \citet{kipping2013parametrizing} beta distribution.

Monte Carlo sampling, while able to explore the parameter space around a best-fit solution, does not deal well with exploring unconstrained parameters with multiple local minima. 
Therefore, in order to allow our model to explore a single solution, we included normal priors on period and $t_0$ using the values and uncertainties from the TOI catalogue in the case of the $2.54$d planet, and from the RV periodogram in the case of the $6.7$d planet.
In all cases, we artificially inflated these uncertainties to make sure the parameters were not over-constrained by their priors, which is confirmed by noting that the posterior distributions are, in all cases, narrower than the priors.

The combined model, built using the \textsf{exoplanet} \citep{exoplanet:exoplanet} package, was sampled using the No-U Turn Sampler (NUTS) in the Hamiltonian Monte Carlo \texttt{PyMC} back-end \citep{exoplanet:pymc3} using 5 independent chains with 2000 steps and an additional 500 steps burn-in.
This produced 10000 independent samples.
Model priors and posteriors are displayed in table \ref{tab:planetparlong}.

\begin{figure*}
	\includegraphics[width=\textwidth, trim={1.45cm 0.2 0.95cm 0.5}]{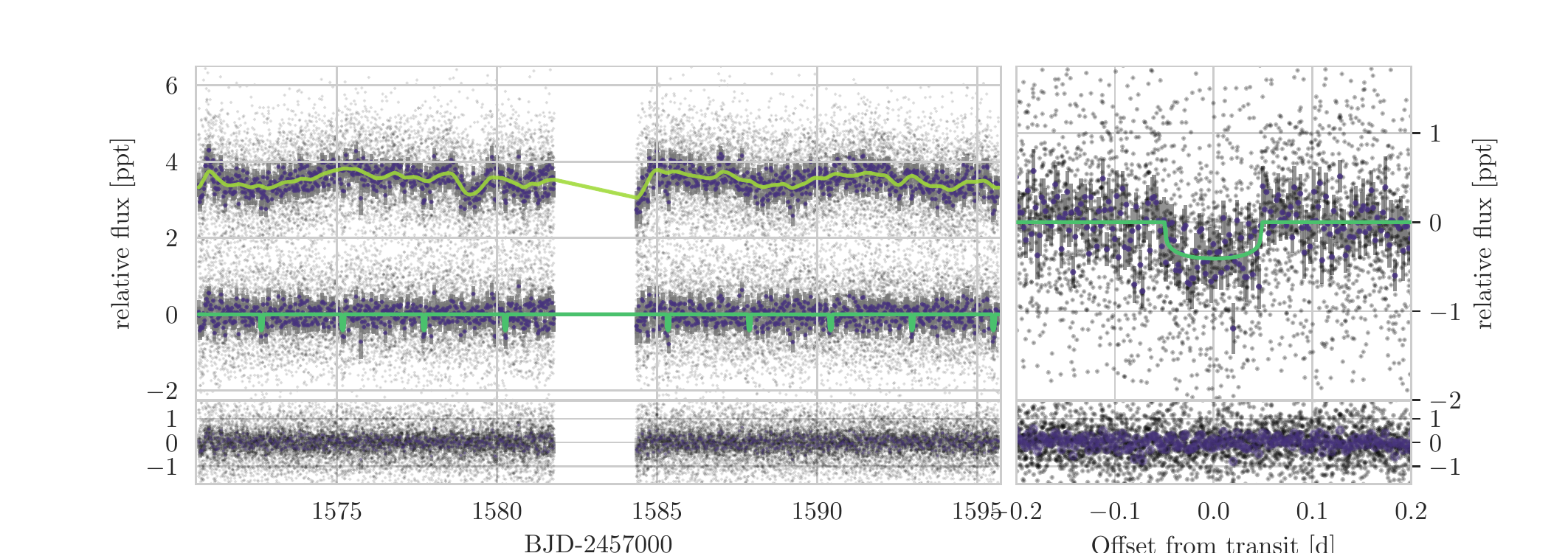}
    \caption{\tess{} photometry, where black dots represent individual 2-minute cadence data and dark circles (with errorbars) represent 30-minute bins. Upper left: \tess{} \texttt{PDC\_SAP} time series with best-fit GP model (both offset by 3.5ppt), and GP-subtracted lightcurve with the best-fit transit model over-plotted (no offset). Lower left: residuals, with both GP model and transit models subtracted from the lightcurve. Upper right: phase-folded lightcurve of \Tplanet{} zoomed to the transit. Lower right: phase-folded residuals. }
    \label{fig:photometry}
\end{figure*}

The results from the combined model are shown in tables \ref{tab:derived_pars} and \ref{tab:planetparlong}, with the \harps{} RV timeseries and best-fit models shown in figure \ref{fig:RVs}, phase-folded RVs and model shown in figure \ref{fig:phase_fold_rvs}, and \tess{} photometry and best-fit light curves shown in figure \ref{fig:photometry}.

\begin{table}
	\centering
        \begin{tabular}{lcc}
        \hline
        \hline
        Parameter & \Tplanet{} & \Tplanetc{}\\
        \hline
        \hline
        Epoch, $t_0$ [BJD-2457000] &  \Ttzerozero{}  &  \Ttzeroone  \\
        Orbital Period, $P$ [d] &  \TPzero{}  &  \TPone{} \\
        Semi-major Axis, $a$ [AU] &  \Tsmazero{}  &  \Tsmaone{}  \\
        Orbital Eccentricity, $e$ &  \Tecczero{}  & \Teccone{}  \\
        Argument of periastron, $\Omega$ &  \Tomegazero{} &  \Tomegaone{}  \\
        Radius ratio [$R_p/R_s$] &  \Tror{}  & --- \\
        Radius, $R_p$ [$R_\oplus$] &  \Trpl{}  & --- \\
        Impact Parameter, b &  \Tb{}  & --- \\
        Transit duration, $t_D$ [d] &  \Ttdurzero{}  & --- \\
        RV semi-amplitude, $K$ [\ms{}] &  \TKzero{}  &  \TKone{}  \\
        Planet Mass, $M_p$ [$M_\oplus$] &  \TMpzero{}  &  \TMpone{}$^{\star}$  \\
        Planet Density, $\rho_p$ [${\rm gcm}^{-3}$] &  \Trhopgcmthree{}  & --- \\
        Insolation, $S$ [${\rm kW m}^{-2}$] &  \TSinzero{} &  \TSinone{} \\
        Surface Temperature, $T_p$ [K]$^{\dagger{}}$ & \TTsurfpzero{}  &  \TTsurfpone{} \\
        \hline
        \hline
        \end{tabular}
	\caption{Derived planet properties. $^{\star}$ The mass of planet c refers to the $M_p\sin{i}$. $^{\dagger{}}$ Surface temperature assumes a uniform surface and an albedo of 0.2.}
	\label{tab:derived_pars}
\end{table}

\section{Discussion}\label{sect:discus}

\subsection{Evidence for \Tplanetc{}}\label{sect:planetc}
The periodogram of the activity-corrected radial velocity timeseries showed a clear signal at $6.75$\,d, even stronger than that of the planet at $2.54$\,d (Figure \ref{fig:rv_decorr}).
No such signal was found by TESS' automatic TPS; however, there is a chance such a signal may have been missed.
A search using the \texttt{transit least squares} algorithm \citep{hippke2019optimized} on the \Tplanet{}-subtracted lightcurve found no signal around 6.7\,d, and a visual inspection of the lightcurve around the likely epochs of transits (given the limits from the RV detection) reveals no candidate dips associated with an outer candidate.
Indeed, when running a combined model of two transiting planets, with constraints on orbits from the RVs, the posteriors for the radius of the outer planet were $<0.64$\,\rearth{} at $1-\sigma$ which, given the \TMpone{}\,\mearth{} mass of \Tplanetc{}, would be physically impossible, even with an iron-core.
Therefore, we come to the conclusion that \Tplanetc{} is likely non-transiting.

In order to assess whether the RV signal alone warrants calling \Tplanetc{} a confirmed planet or merely a candidate, we ran two combined models with identical priors and with one model including a non-transiting planet around $6.7$d.
We then burned-in each model for 500 samples and ran the \texttt{find\_MAP} function in \texttt{PyMC3} to find the maximum likelihood for each model, allowing us to compare the difference in Bayesian Information Criterion ($\Delta{\rm BIC}$) between the models.
The resulting value of $\Delta{\rm BIC}= $\,\TdeltaBIC{} clearly favours a two-planet model over a single planet model, with $\Delta{\rm BIC}>10$ suggesting "Very Strong" evidence over the null hypothesis. 

\begin{figure}
	\includegraphics[width=\columnwidth, trim={0.2cm 0.6cm 0.9cm 0.6cm}]{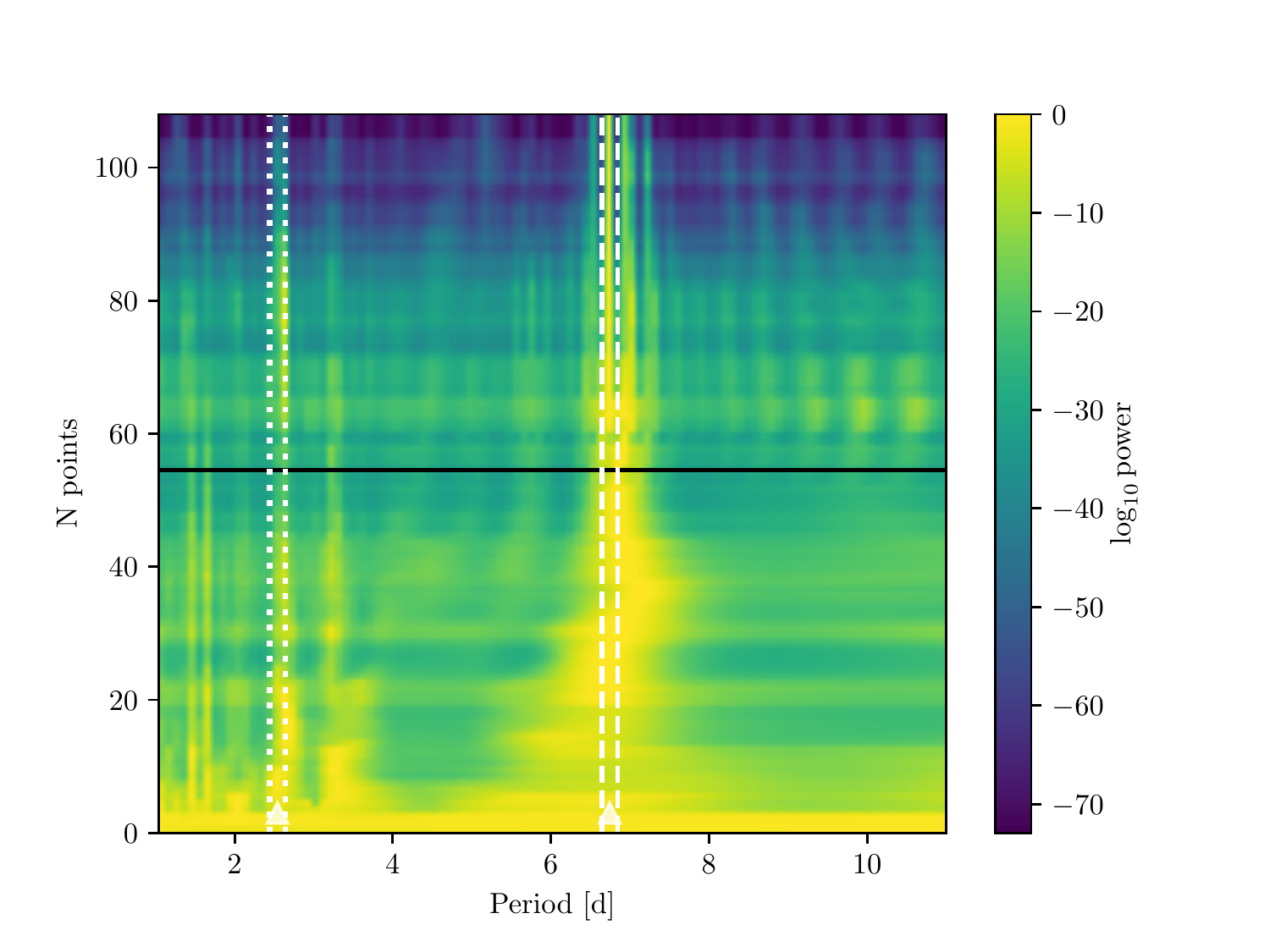}
    \caption{A 2D BGLS periodogram of \Tstar{} radial velocities (after decorrelation and subtraction of the strongest rotation signal) performed on increasing numbers of radial velocity points, as proposed by \citet{mortier_2017_bgls2}. Periods which maintain signal-strength and periodicity as a function of observation number suggest coherent (and therefore planetary) signals. The two white vertical bands show our modelled periods of each planet. The vertical bands seen are due to signal aliases in the second observing season due to the data gap, which is marked with a black horizontal line. \label{fig:RVSignalCoherence}
    }
\end{figure}

Another test for the RV signal of \Tplanetc{} is the coherence of the signal over time, as radial velocity variation due to, e.g., stellar variability is not likely to remain coherent over multiple observing seasons.
We verified this two ways using the decorrelated and rotation-subtracted radial velocities previously used to form RV periodograms (see Figure \ref{fig:rv_decorr}).
First we processed each season individually, finding that the signals at 2.451\,d and 6.75\,d coincide with peaks during both seasons, albeit at lower signal strength.
Next we applied the Bayesian generalised Lomb-Scargle periodogram \citep[BGLS,][]{mortier_2015_bgls} to subsets of our RV time series to test signal coherence as per the technique of \citet{mortier_2017_bgls2}.
Figure \ref{fig:RVSignalCoherence} shows that the signal of \Tplanetc{} passes this test - remaining evident even in datasets with only a handful of datapoints.

It should be noted that the period of \Tplanetc{}, at \TPone{}\,d, is close to the $P_{\rm rot}/3$ harmonic.
However, there appears little evidence of a signal in the RV periodogram at $P_{\rm rot}/2$, so a large coherent signal at $P_{\rm rot}/3$ would be unexpected.
However, it is possible that with certain inclinations and spot locations such harmonics may be boosted \citep{vanderburg2016radial,boisse2011disentangling}.
Interestingly the periodogram of the S-index data does show a strong peak at $P_{\rm rot}/2$ and a weaker peak at $P_{\rm rot}/3$, but this occurs at $7.25$\,d---significantly separated from the RV peak at \TPone{}.
While we confirm the presence of this second planet, as given $\Delta{\rm BIC}>10$, the amplitude of the signal may be affected by the presence of a signal at $P_{\rm rot}/3$, therefore the mass of \Tplanetc{} should be treated as uncertain. 

Multiple lines of evidence point to the signal of \Tplanetc{} being planetary in origin.
Future RV measurements should help further disentangle stellar rotation and the signal amplitude, and may even reveal new candidates in this system.

The majority of short-period multi-planet systems are typically aligned with mutual inclinations of only a few degrees \citep{lissauer2011architecture, 2012A&A...541A.139F,winn2015occurrence}.
To investigate whether this could also be true for \Tstar{}, we used the derived impact parameter of planet b and the semi-major axis ratio of b \& c to calculate the expected impact parameter of planet c in a perfectly co-planar scenario ($b_c = 0.60\pm0.42$) and the minimum mutual inclination ($\Delta{}i = 1.6^{+1.4 \circ}_{-1.6}$).
Therefore, the \Tstar{} planetary system is still consistent with an aligned planetary system.

Throughout this work we quote $M_p\sin{i}$ for \Tplanetc{}. 
However, a clear non-detection of transits can constrain a planet's inclination, and therefore also reduce the lower limit on a planet's mass.
However, in this case, the reduction in minimum mass caused by assuming $b>1.0$ is smaller than 0.25\%. Therefore, including this factor would not significantly change the mass estimate from $M_p\sin{i}$.
It is also worth noting that planets b \& c have an orbital period ratio near 8/3, although harmonics beyond $2:1$ are highly unlikely to create measurable TTVs \citep{deck2015measurement}.

\subsection{A solar analogue?}\label{sect:solaranal}
It is remarkable to note just how sun-like \Tstar{} is, with a radius, \teff{} and \logg{} all within 1-sigma uncertainties of solar values, with the exception of its slightly higher metallicity (\feh{}$ = 0.14\pm0.02$), and correspondingly lower C \& O (see Table \ref{tab:abunds}) \citep[e.g.][]{Franchini2020,2015A&A...576A..89B}.
We speculate that the higher metallicity may explain why \Tstar{} was able to form close-in mini-Neptunes \citep{mulders2016super,bitsch2020influence}, which do not exist in our solar system.

\Tstar{} is also nearly the same age as the Sun, as can be seen in both the Yttrium-based ages (Table \ref{tab:abunds}), and from the rotation rate ($\sim$22\,d from archival photometry, spectroscopy timeseries, \& \vsini{}).
Indeed, this rotation rate is marginally faster than the Sun \citep[25--26.5\,d when measured with HARPS-N and converted to sidereal period,][]{milbourne2019harps}.
This could be explained by the fact that \Tstar{} is slightly younger, the Sun rotates slower than average \citep{2008ApJ...684..691R}, or the presence of short-period planets has tidally inhibited the slow-down of \Tstar{}, although the effect for such small planets is likely to be small \citep{bolmont2012effect}.

Thanks to their similarities, \Tstar{} and its planets could prove a useful comparison to the Sun and the solar system in the future.

\subsection{Composition}\label{sect:internal}

\begin{figure}
	\includegraphics[width=\columnwidth]{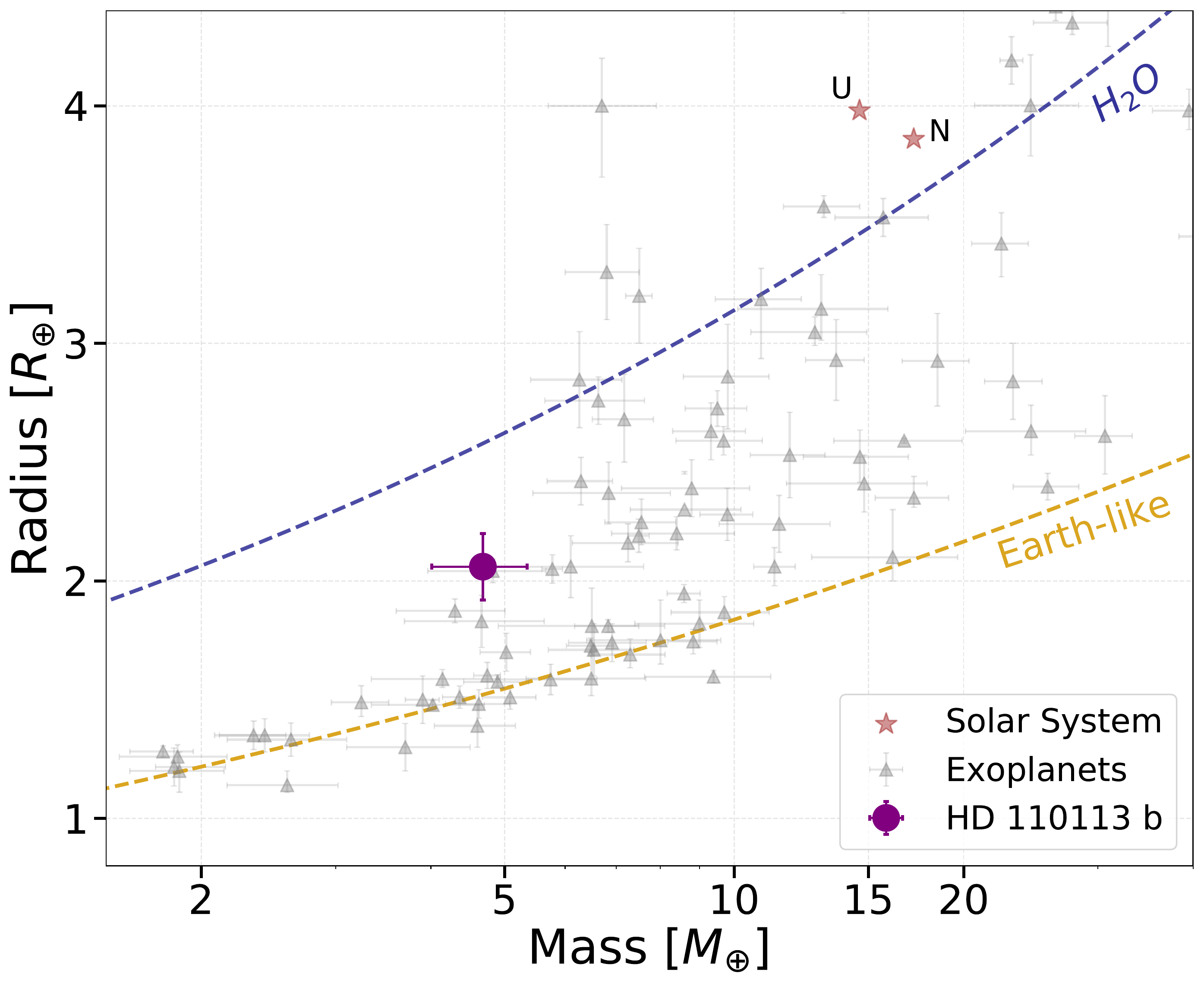}
    \caption{Mass-radius diagram of exoplanets with accurate mass and radius determination (Otegi et al. 2019). Also shown are the mass-radius relations for Earth-like and pure water compositions.\label{fig:MR_Diagram}
    }
\end{figure}

To explore the composition of \Tplanet{}, we performed 4-layer interior structure modelling, using as inputs the mass and radius determined by our joint modelling of \tess{} photometry and \harps{} RVs.
We followed the method of \citet{otegi2020revisited}, which assumes a pure iron core, a silicate mantle, a non-gaseous water layer, and a H-He atmosphere.
In order to quantify the degeneracy between the different interior parameters and produce posterior probability distributions, we use a generalized Bayesian inference analysis with a Nested Sampling scheme \citep[e.g.][]{2014A&A...564A.125B}.
The interior parameters that are inferred include the masses of the pure-iron core, silicate mantle, water layer and H-He atmosphere.
The ratios of Fe/Si and Mg/Si found in stars is expected to be mirrored in the protoplanetary material, and therefore in the internal structures of exoplanets \citep{dorn2015}.
Hence, we use the values found by our stellar abundance analysis as a proxy for the core-to-mantle ratio.
Given the observed molar ratio of Fe/Si ($1.08\pm0.07$, Table \ref{tab:abunds}) is higher than that of the Sun  \citep[0.85,][]{lodders2009}, we would expect planetary material around \Tstar{} to be more iron-rich than Earth.

Table \ref{tab:interior} lists the inferred mass fractions of the core, mantle, water-layer and H-He atmosphere from the interior models.
Due to the nature of the measurements, interior models cannot distinguish between water and H-He as the source of low-density material. 
Therefore, we ran both a 4-layer model and two 3-layer models, which leave out the H$_{2}$O and H-He envelopes, respectively.
In the case of a H-He envelope, we find that the planet is only $\sim1$\% H-He by mass, with an iron-rich rocky interior making up 99\% of the planet.
Any water present would likely decrease the core, mantle \& gaseous envelope fractions.
However, a gas-free model would require $73 ^{+10} _{-13}$\% water.
Such a high water-to-rock ratio is challenging from formation point of view.
Therefore \Tplanet{} almost certainly has a significant gaseous envelope.
Stars with super-solar metallicities are also less likely to host water-rich planets due to a higher C/O ratio \citep{bitsch2020influence}, making a water-rich composition even less likely.

Figure \ref{fig:MR_Diagram} shows the mass radius relation (M-R relation) for Earth-like and pure water compositions (where the pure water line corresponds to a surface pressure of 1 bar, and without a water-vapor atmosphere).
Also shown are exoplanets with accurate mass and radius determinations from \citet{otegi2020revisited}.
The position of \Tplanet{} makes it one of the lowest-density worlds found with $M_p<5$\,\mearth{}, and among a small class of low-density low-mass planets which includes $\pi{}$ Men c \citep{huang2018tess} and GJ 9827 b \citep{niraula2017three}.

\begin{table}
\caption{Inferred interior structure properties of TOI-755b.}
\label{tab:interior}
\begin{tabular}{lccc}
\hline
\hline
Constituent & With H-He [\%] & With H$_{2}$O [\%] & 4-layer [\%] \\
\hline
\hline
$M_{\rm core}/M_{\rm total}$ & $47 ^{+26} _{-24} $ & $8 ^{+7} _{-6}$ & $25 ^{+28} _{-18}$\\
$M_{\rm mantle}/M_{\rm total}$ & $ 53 ^{+23} _{-24} $ & $17 ^{+11} _{-9}$ & $36 ^{+31} _{-19}$\\
$M_{\rm water}/M_{\rm total}$ & --- & $73 ^{+10} _{-13}$ & $38 ^{+31} _{-24}$\\
$M_{\rm H-He}/M_{\rm total}$ & $1.0 ^{+0.3} _{-0.5}$ & --- & $0.102  ^{+0.04} _{-0.03}$\\
\hline
\hline
\end{tabular}
\end{table}

\begin{figure}
	\includegraphics[width=\columnwidth, trim={0.8cm 0.5cm 1.05cm 0.2cm}]{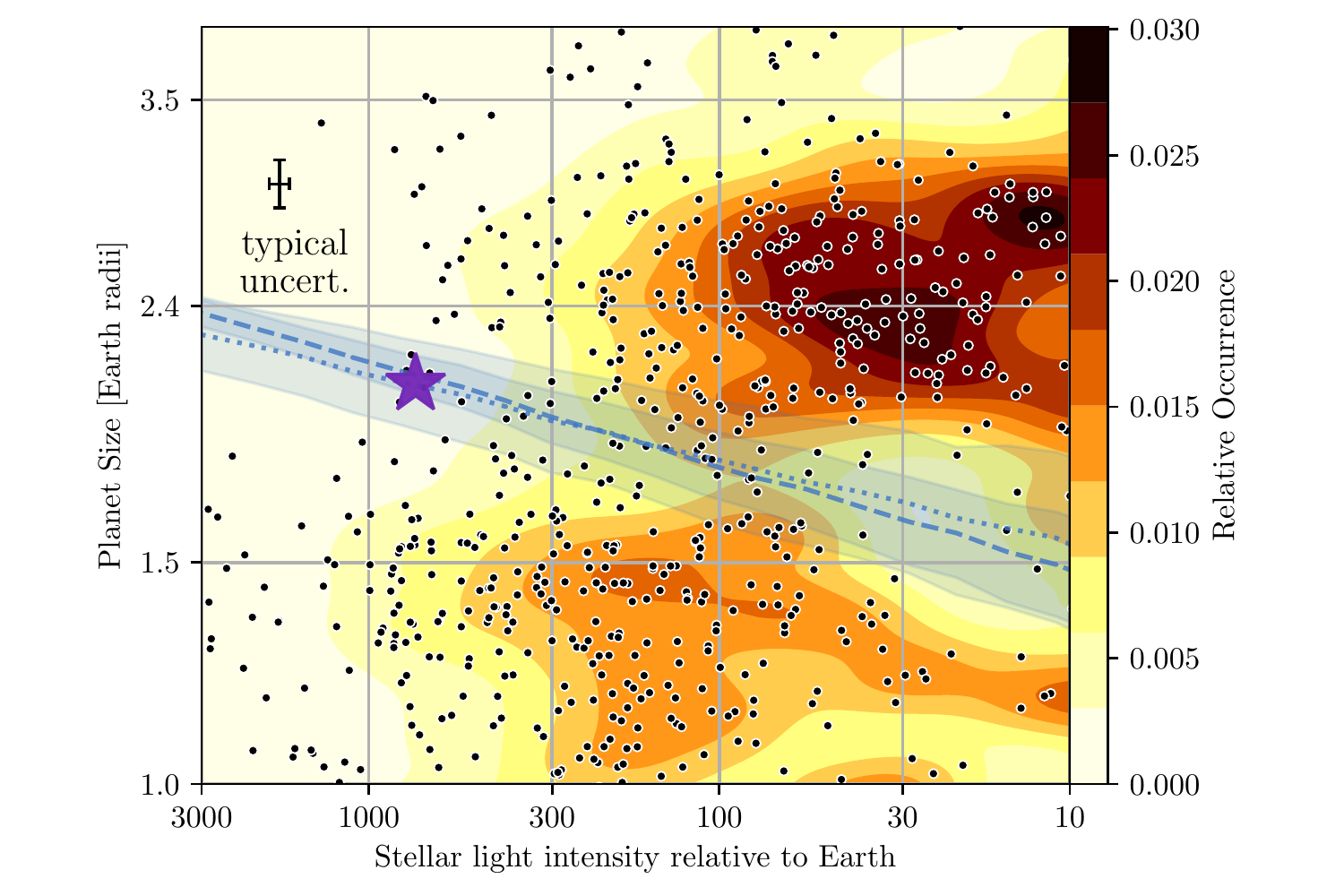}
    \caption{The distribution of Kepler planets by both insolation \& planetary radius plot, with underlying occurrence distributions adapted from \citep{martinez2019spectroscopic}. \Tplanet{} is included as a purple star. The best-fit positions of radius valleys from both \citet{martinez2019spectroscopic} (dashed) and \citet{van2018asteroseismic} (dotted) are plotted in blue, with conversion from period to insolation performed using the average stellar parameters in the Kepler samples. Typical uncertainties for both \Tplanetc{} and for the Kepler sample are shown in the top left.\label{fig:insolation}
    }
\end{figure}

\subsection{Evaporation}\label{sect:evap}
With an insolation of \TSinzero{}\,kW\,m$^{-2}$ ($\sim736\,S_\oplus$), it is extremely likely that \Tplanet{} has been moulded by strong stellar radiation in some way.
This is further suggested by placing \Tplanet{} on the insolation-radius plots of \citet{fulton2017california} and \citet{martinez2019spectroscopic}, which clearly show the "radius valley" (see Fig. \ref{fig:insolation}.
The negative slope of the valley with insolation means that, even with a radius of \Trpl{}\,\rearth{}, \Tplanet{} is positioned exactly within it.

Using both rotation and age, we predict a current X-ray luminosity ($L_x/L_{\rm bol}$) of between $8.5\times10^{-7}$ \citep[with Prot;][]{wright2018stellar} and $2.74\times10^{-6}$ \citep[with age;][]{jackson2012coronal}.
This implies total X-ray luminosities on the order of $3.3\times10^{27}$ to $2.7\times10^{28}$ erg and mass-loss rates (assuming an energy-limited regime) of between $5\times10^{9}$ and $9\times10^{9}$\,gs$^{-1}$ (0.026-- 0.05\,$M_\oplus {\rm Gyr}^{-1}$).
This is comparable to both GJ 436 b and Pi Men c under similar assumptions \citep{king2019xuv}.
Therefore, while it is currently highly irradiated, \Tplanet{} is unlikely to currently be losing large quantities of its H-He atmosphere to space.

However, the integrated sum of mass-loss since the planet's formation is substantial, as young stars are typically far more active and far more X-ray luminous.
We calculate that, assuming the current mass and radius, as much as 10\% of the planet's mass may have been lost through evaporation.
The models of \citet{zeng2019growth} suggest that a 1000\,K planet with $>5$\% hydrogren and a $5.25M_\oplus$ core would have been $>8.5$\,\rearth{} in radius, suggesting that \Tplanet{} potentially started as an extremely low-density Jupiter-radius world which was quickly stripped.
How such a low-mass world came to possess such a large gaseous atmosphere raises more questions.

In any case, it is highly likely that \Tplanet{} started with a thicker atmosphere of H-He, which, due to both evaporative and core-powered mass-loss, it lost much of over time.
However, this is typically a runaway process: planets which lose the majority of their gas (i.e. those in the radius valley) typically lose it all \citep{owen2017evaporation}.
Therefore the main unanswered question is: how did \Tplanet{} escape becoming a naked core devoid of volatile envelope?
Here we propose two solutions to this:

1) \Tplanet{} started with a large envelope of H-He, perhaps as much as 10\%, which was gradually lost to evaporation and core-powered heating over time. 
But it had just enough gas to walk the tight-rope between keeping hold of a thick atmosphere and being completely stripped such that, at the point that evaporative forcing stopped, \Tplanet{} still had $\sim1\%$ of H-He by mass.
The models of \citet[][Figure 4]{rogers2020unveiling} suggest such a scenario is possible and may occur for planets that start gas-rich with around 4\% H-He.

2) \Tplanet{} did lose almost all of its H-He to evaporation and core-powered mass-loss.
The current density is therefore explained by the planet having a large water content (e.g. an icy core), with potential out-gassing of a He-depleted secondary atmosphere contributing to the extended radius.
Indeed, our composition calculations include water in only solid \& liquid states; therefore a thick steam \citep[or supercritical;][]{mousis2020irradiated} H$_{2}$O atmosphere could reduce the density without requiring $>50\%$ H$_{2}$O.

One final solution might be that \Tplanet{} and \Tplanetc{} underwent late-stage migration to their current positions, thereby avoiding much of the evaporative forcing caused by the X-ray emissions of the young star. 
However, there is no theoretical mechanism in which a low-eccentricity 2-planet system could undergo such late-stage migration long after the dispersal of the protoplanetary disc.
Instead, multi-planet systems are capabale of undergoing early-stage migration damped by the protoplanetary gas disc \citep{2006A&A...450..833C,2019MNRAS.486.3874C}, and massive single planets are thought capable of undergoing late-stage, high-eccentricity scattering onto shorter orbits \citep{2008ApJ...686..621F,2012ApJ...751..119B}.
We therefore consider a solution through in-situ processes more plausible than through migration.

These two predictions may be testable with future transmission spectroscopy observations, e.g. with \textit{JWST} \citep{2016ApJ...817...17G,2014PASP..126.1134B}.

\subsection{Potential for future observations}\label{sect:future}
The low-density nature of this hot mini-Neptune, combined with its bright host star, may enable transmission spectroscopy observations.
Such measurement could test the hypotheses noted above, and search for a low-molecular weight primary atmosphere dominated by H-He, or a high molecular weight secondary atmosphere dominated by an overabundance of water vapour \citep{2020arXiv201011867B}.
To test this, we computed the emission and transmission spectroscopy metrics from \citet{kempton2018framework}.

We find that, amongst small planets with $R_p<4R_\oplus$ \citep{akeson2013nasa}\footnote{\url{https://exoplanetarchive.ipac.caltech.edu/cgi-bin/nstedAPI/nph-nstedAPI?table=exoplanets&select=*&format=csv}, accessed 2020-Oct-18}, \Tplanet{} ranks in the top 3\% most amenable for emission and the top 5\% for transmission spectroscopy with \textit{JWST}.
Although, when compared to one of the most favourable \textit{JWST} targets: the low-density mini-Neptune GJ\,1214\,b, \Tplanet{} provides only around 10\% the SNR in both transmission \& emission --- as is expected when comparing with a planet whose transits are 36 times deeper.

\Tplanet{} will be also re-observed by \tess{} during Sector 37\footnote{\url{https://heasarc.gsfc.nasa.gov/cgi-bin/tess/webtess/wtv.py?Entry=73228647}}, and could also be observed by ESA's \textit{CHEOPS} telescope \citep{benz2020cheops}, both of which would improve the radius precision below the currently measured value of 7\%, thereby improving our knowledge of the internal structure of \Tplanet{}.

\section{Conclusion}
\label{sec:conclusion}
We have presented the detection and confirmation of \Tplanet{}, which was initially spotted as \TTplanet{} in \tess{} with an SNR of only $7.6\sigma$ and transit depth of \Tdepth{}.
This marks one of the lowest-SNR signals yet to be confirmed from \tess{}, and is testament to the unique ability of \tess{} to find planet candidates around bright stars which can be redetected and characterised through independent RV campaigns.

High-resolution imaging and ground-based photometry rules out the presence of nearby companions and potential nearby eclipsing binaries, thereby limiting the number of false-positives and giving us confidence to follow such a low-SNR signal.
Our subsequent HARPS campaign obtained more than 100 \harps{} spectra in order to characterise both \Tplanet{} and its bright ($G=9.9$\,mag) star.

Analysis of these spectra revealed \Tstar{} to be a Sun-like G-type star with slightly super-solar metallicity, but solar \teff{}, \logg{} and age.
The RV timeseries also revealed strong activity on \Tstar{} with a rotation period of \Tperiod{}\,d---a timescale corroborated by archival WASP photometry.

Removing this rotation period using both linear decorrelation and a co-fitted GP using S-index and FWHM activity indicators revealed the presence of two Keplerian signals, at \TPzero{}\,d and \TPone{}\,d.
The inner signal, from a planet with mass \TMpzero{}\,\mearth{}, corresponded to the detected \tess{} candidate with a radius, as modelled from the \tess{} photometry, of \Trpl{}\,\rearth{}.
The outer signal, from a planet with $M_p\sin{i}$ of \TMpone{}\,\mearth{} did not correspond to any transit events in the \tess{} lightcurve, and therefore is likely non-transiting.
We were able to confirm it as a planet through Bayesian model comparison which showed $\Delta{\rm BIC}= $\TdeltaBIC{} in favour of a 2-planet model.

The estimated density of \Tplanet{} is \Trhopgcmthree{}\,\gcm{}---far lower than would be expected from a rocky core.
By modelling four potential constituents---an iron core, silicate mantle, water ocean and H-He atmosphere---we were able to rule out a gasless composition for \Tplanet{}, suggesting that it has between 0.07 and 1.5\% H-He by mass.
This is surprising given \Tplanet{}'s position in the "radius valley" between gaseous mini-Neptunes and rocky super-Earths, and we suggest two possibilities for this unexpectedly low density: either \Tplanet{} has a water-rich core and secondary atmosphere, or it began with a thick H-He envelope and managed to retain a small fraction of it despite significant evaporation and/or heating.
Follow-up spectroscopy observations with the next generation of telescopes may reveal the answer, as well as far more about this interesting system.



\section*{Acknowledgements}
{\footnotesize We thank Rapha{\"e}lle Haywood, Maximillian G{\"u}nther and Francois Bouchy for discussion on disentangling RV activity from signals.\\
This paper includes data collected by the \tess{} mission.
Funding for the \tess{} mission is provided by the NASA Explorer Program and NASA's Science Mission directorate.
We acknowledge the use of public \tess{} Alert data from pipelines at the \tess{} Science Office and at the \tess{} Science Processing Operations Center. 
This paper includes data collected by the \tess{} mission, which are publicly available from the Mikulski Archive for Space Telescopes (MAST).\\
This research has made use of the Exoplanet Follow-up Observation Program website, which is operated by the California Institute of Technology, under contract with the National Aeronautics and Space Administration under the Exoplanet Exploration Program. \\
Resources supporting this work were provided by the NASA High-End Computing (HEC) Program through the NASA Advanced Supercomputing (NAS) Division at Ames Research Center for the production of the SPOC data products. \\
This study is based on observations collected at the European Southern Observatory under ESO programme 1102.C-0249.\\ 
We thank the Swiss National Science Foundation (SNSF) and the Geneva University for their continuous support to our planet search programs. This work has been in particular carried out in the frame of the National Centre for Competence in Research {\it PlanetS} supported by the Swiss National Science Foundation (SNSF).\\ 
This publication makes use of The Data \& Analysis Center for Exoplanets (DACE), which is a facility based at the University of Geneva (CH) dedicated to extrasolar planets data visualisation, exchange and analysis. DACE is a platform of the Swiss National Centre of Competence in Research (NCCR) PlanetS, federating the Swiss expertise in Exoplanet research. The DACE platform is available at \url{https://dace.unige.ch}. \\ 
This work makes use of observations from the LCOGT network.\\ 
(Some of the) Observations in the paper made use of the High-Resolution Imaging instrument(s) ‘Alopeke (and/or Zorro). ‘Alopeke (and/or Zorro) was funded by the NASA Exoplanet Exploration Program and built at the NASA Ames Research Center by Steve B. Howell, Nic Scott, Elliott P. Horch, and Emmett Quigley. Data were reduced using a software pipeline originally written by Elliott Horch and Mark Everett. ‘Alopeke (and/or Zorro) was mounted on the Gemini North (and/or South) telescope of the international Gemini Observatory, a program of NSF’s OIR Lab, which is managed by the Association of Universities for Research in Astronomy (AURA) under a cooperative agreement with the National Science Foundation. on behalf of the Gemini partnership: the National Science Foundation (United States), National Research Council (Canada), Agencia Nacional de Investigación y Desarrollo (Chile), Ministerio de Ciencia, Tecnología e Innovación (Argentina), Ministério da Ciência, Tecnologia, Inovações e Comunicações (Brazil), and Korea Astronomy and Space Science Institute (Republic of Korea).\\ 
Based in part on observations obtained at the Southern Astrophysical Research (SOAR) telescope, which is a joint project of the Minist\'{e}rio da Ci\^{e}ncia, Tecnologia e Inova\c{c}\~{o}es (MCTI/LNA) do Brasil, the US National Science Foundation’s NOIRLab, the University of North Carolina at Chapel Hill (UNC), and Michigan State University (MSU), and the international Gemini Observatory, a program of NSF’s NOIRLab, which is managed by the Association of Universities for Research in Astronomy (AURA) under a cooperative agreement with the National Science Foundation. on behalf of the Gemini Observatory partnership: the National Science Foundation (United States), National Research Council (Canada), Agencia Nacional de Investigaci\'{o}n y Desarrollo (Chile), Ministerio de Ciencia, Tecnolog\'{i}a e Innovaci\'{o}n (Argentina), Minist\'{e}rio da Ci\^{e}ncia, Tecnologia, Inova\c{c}\~{o}es e Comunica\c{c}\~{o}es (Brazil), and Korea Astronomy and Space Science Institute (Republic of Korea).\\ 
\\
H.P.O. acknowledges support from NCCR/Planet-S via the CHESS fellowship.\\
D.J.A. acknowledges support from the STFC via an Ernest Rutherford Fellowship (ST/R00384X/1).\\ 
V.A., E.D.M., N.C.S., O.D.S.D. \& S.C.C.B. acknowledge support by FCT - Funda\c{c}\~ao para a Ci\^encia e a Tecnologia (Portugal) through national funds and by FEDER through COMPETE2020 - Programa Operacional Competitividade e Internacionaliza\c{c}\~ao by these grants: UID/FIS/04434/2019; UIDB/04434/2020; UIDP/04434/2020; PTDC/FIS-AST/32113/2017 \& POCI-01-0145-FEDER-032113; PTDC/FIS-AST/28953/2017 \& POCI-01-0145-FEDER-028953.\\ 
V.A. and E.D.M. further acknowledge the support from FCT through Investigador FCT contracts IF/00650/2015/CP1273/CT0001 and IF/00849/2015/CP1273/CT0003.\\ 
O.D.S.D. and S.C.C.B. are supported through Investigador contract (DL 57/2016/CP1364/CT0004) funded by FCT.\\ 
J.L-B. is supported by the Spanish State Research Agency (AEI) Projects No.ESP2017-87676-C5-1-R and No. MDM-2017-0737 Unidad de Excelencia "Mar\'ia de Maeztu"- Centro de Astrobiolog\'ia (INTA-CSIC)\\ 
D.D. acknowledges support from the TESS Guest Investigator Program grant 80NSSC19K1727 and NASA Exoplanet Research Program grant 18-2XRP18\_2-0136.\\ 
B.V.R. thanks the Heising-Simons Foundation for support.\\ 
T.D. acknowledges support from MIT's Kavli Institute as a Kavli postdoctoral fellow\\ 
A.O. acknowledges support from an STFC studentship.\\ 
S.H. acknowledges CNES funding through the grant 837319\\ 
D.J.A.B. acknowledges support by the UK Space Agency.\\ 

This research made use of the following python software:
\texttt{exoplanet} \citep{exoplanet:exoplanet} and its dependencies \citep{exoplanet:agol19, exoplanet:astropy13, exoplanet:astropy18, exoplanet:exoplanet, exoplanet:foremanmackey17, exoplanet:foremanmackey18, exoplanet:luger18, exoplanet:pymc3, exoplanet:theano};
\texttt{numpy} \citep{harris2020array}; 
\texttt{scipy} \citep{virtanen2020scipy}; 
\texttt{pandas} \citep{mckinney2011pandas}; 
\texttt{astropy} \citep{robitaille2013astropy}; 
\texttt{matplotlib} \citep{hunter2007matplotlib};
\texttt{AstroImageJ} \citep{Collins:2017},;
\texttt{TAPIR} \citep{Jensen:2013}.
}


\section{Data availability}
The data underlying this article is publicly available - \tess{} data is stored on the Mikulski Archive for Space Telescopes (MAST) at \url{https://archive.stsci.edu/tess/}, while \harps{} data is both available on the Data \& Analysis Center for Exoplanets (DACE) at \url{https://dace.unige.ch/}, and in Appendix tables \ref{tab:spec1} \& \ref{tab:spec2}.

\bibliographystyle{mnras}
\bibliography{example} 




\appendix
\section{Author Affiliations}\label{sec:affiliations}
\textsuperscript{\hypertarget{affil_1}{1}} NCCR/PlanetS, Centre for Space \& Habitability, University of Bern, Bern, Switzerland\\
\textsuperscript{\hypertarget{affil_2}{2}} Department of Physics and Kavli Institute for Astrophysics and Space Research, Massachusetts Institute of Technology, Cambridge, MA 02139, USA\\
\textsuperscript{\hypertarget{affil_3}{3}} Centre for Exoplanets and Habitability, University of Warwick, Gibbet Hill Road, Coventry, CV4 7AL, UK\\
\textsuperscript{\hypertarget{affil_4}{4}} Department of Physics, University of Warwick, Gibbet Hill Road, Coventry CV4 7AL, UK\\
\textsuperscript{\hypertarget{affil_5}{5}} Instituto de Astrof\'isica e Ci\^encias do Espa\c{c}o, Universidade do Porto, CAUP, Rua das Estrelas, 4150-762 Porto, Portugal\\
\textsuperscript{\hypertarget{affil_6}{6}} Harvard-Smithsonian Center for Astrophysics, 60 Garden St, Cambridge, MA, 02138, USA\\
\textsuperscript{\hypertarget{affil_7}{7}} NASA Ames Research Center,Moffett Field, CA 94035, USA\\
\textsuperscript{\hypertarget{affil_8}{8}} Astrophysics Group, Keele University, Staffs ST5 5BG, U.K. \\
\textsuperscript{\hypertarget{affil_9}{9}} Centro de Astrobiologi\'ia (CAB,CSIC-INTA), Dep. de Astrof\'isica, ESAC campus, 28692, Villanueva de la Ca\~nada, Madrid, Spain\\
\textsuperscript{\hypertarget{affil_10}{10}} Geneva Observatory, University of Geneva, Chemin des Mailettes 51, 1290 Versoix, Switzerland\\
\textsuperscript{\hypertarget{affil_11}{11}} Departamento de F\'isica e Astronomia, Faculdade de Ci\^{e}ncias, Universidade do Porto, Rua do Campo Alegre, 4169-007 Porto, Portugal\\
\textsuperscript{\hypertarget{affil_12}{12}} Dunlap Institute for Astronomy and Astrophysics, University of Toronto, 50 St. George Street, Toronto, Ontario M5S 3H4, Canada\\
\textsuperscript{\hypertarget{affil_13}{13}} Cerro Tololo Inter-American Observatory, Casilla 603, La Serena, Chile\\
\textsuperscript{\hypertarget{affil_14}{14}} SETI Institute\\
\textsuperscript{\hypertarget{affil_15}{15}} American Association of Variable Star Observers, 49 Bay State Road, Cambridge, MA 02138, USA\\
\textsuperscript{\hypertarget{affil_16}{16}} International Center for Advanced Studies (ICAS) and ICIFI (CONICET), ECyT-UNSAM, Campus Miguelete, 25 de Mayo y Francia, (1650) Buenos Aires, Argentina.\\
\textsuperscript{\hypertarget{affil_17}{17}} Department of Physics and Astronomy, University of New Mexico, 1919 Lomas Blvd NE, Albuquerque, NM 87131, USA\\
\textsuperscript{\hypertarget{affil_18}{18}} Aix Marseille Univ, CNRS, CNES, LAM, Marseille, France\\
\textsuperscript{\hypertarget{affil_19}{19}} Institute for Computational Science, University of Zurich,Winterthurerstr. 190, CH-8057 Zurich, Switzerland\\
\textsuperscript{\hypertarget{affil_20}{20}} Kavli Fellow\\
\textsuperscript{\hypertarget{affil_21}{21}} European Southern Observatory, Alonso de Cordova 3107, Vitacura, Santiago, Chile\\
\textsuperscript{\hypertarget{affil_22}{22}} Department of Physics \& Astronomy, Swarthmore College, Swarthmore PA 19081, USA\\
\textsuperscript{\hypertarget{affil_23}{23}} Department of Physics and Astronomy, The University of North Carolina at Chapel Hill, Chapel Hill, NC 27599-3255, USA\\
\textsuperscript{\hypertarget{affil_24}{24}} Department of Astronomy, University of Maryland, College Park, MD 20742, USA\\
\textsuperscript{\hypertarget{affil_25}{25}} Space Telescope Science Institute, 3700 San Martin Drive, Baltimore, MD 21218, USA\\
\textsuperscript{\hypertarget{affil_26}{26}} Department of Earth, Atmospheric and Planetary Sciences, Massachusetts Institute of Technology, Cambridge, MA 02139, USA\\
\textsuperscript{\hypertarget{affil_27}{27}} Department of Physics \& Astronomy, Vanderbilt University, 6301 Stevenson Center Ln., Nashville, TN 37235, USA\\
\textsuperscript{\hypertarget{affil_28}{28}} Department of Astrophysical Sciences, Princeton University, 4 Ivy Lane, Princeton, NJ 08544, USA\\

\section{Extra tables}

\centering
\begin{table*}
\caption{List of free parameters used in the \texttt{exoplanet} combined analysis of the \tess{} light curve and \harps{} radial velocities with their associated prior and posterior distributions.}
\label{tab:planetparlong}
\begin{center}
\begin{tabular}{lcc}
\hline
\hline
Parameter & Prior & Posterior\\
\hline
\hline
\multicolumn{3}{l}{\it Stellar parameters}\\
Stellar surface temperature, \teff{} [K] &  $\mathcal{N}(5732.0,50.0)$  & \TTeff{} \\
Stellar Mass, $M_s$ [$M_{\odot}$] &  $\mathcal{N}(0.9968,0.06)$  & \Tmstar{} \\
Stellar Radius, $R_s$ [$R_{\odot}$] &  $\mathcal{N}(0.968,0.018)$  & \Trstar{} \\
\hline
\multicolumn{3}{l}{\it Orbital parameters}\\
Transit Epoch, $t_0$ [BJD-2457000] b &  $\mathcal{N}(1570.10189,0.1)$  &  \Ttzerozero{} \\
Transit Epoch, $t_0$ [BJD-2457000] c &  $\mathcal{N}(1798.1334,1.0)$  & \Ttzeroone{} \\
Orbital Period, $P$ [d] b &  $\mathcal{N}_{\mathcal{U}}(2.540455,0.002124,2.35,2.6)$  & \TPzero{} \\
Orbital Period, $P$ [d] c &  $\mathcal{N}_{\mathcal{U}}(6.7285,0.05951,6.65,6.8)$  & \TPone{} \\
Orbital Eccentricity, $e$ b &  $\beta(0.867;3.03)^{a}$  & \Tecczero{} \\
Orbital Eccentricity, $e$ c &  $\beta(0.867;3.03)^{a}$  & \Teccone{} \\
Argument of periastron, $\Omega$ b &  $\mathcal{U}(-\pi,\pi)^{b}$  &  \Tomegazero{} \\
Argument of periastron, $\Omega$ c &  $\mathcal{U}(-\pi,\pi)^{b}$  & \Tomegaone{} \\
\hline
\multicolumn{3}{l}{\it Photometric parameters}\\
log radius ratio [$\log{R_p/R_s}$] b &  $\mathcal{U}(-11.513,-2.3023)$  & \Tlogror{} \\
Transit Impact Parameter b & $\mathcal{U}(0,1+R_p/R_s)^{c}$  & \Tb{} \\
Quadratic Limb Darkening $a_{\rm LD}$ &  $\mathcal{N}_{\mathcal{U}}(0.367,0.1,0.0,1.0)$  & \Tustartesszero{}\\
Quadratic Limb Darkening $b_{\rm LD}$ &  $\mathcal{N}_{\mathcal{U}}(0.21,0.1,0.0,1.0)$  & \Tustartessone{} \\
Photometric jitter [$\log{\rm ppt}$] &  $\mathcal{N}(0.7294,5.0)$  &  \Tphotlogerrcontr{} \\
Photometric GP power & $\mathcal{I}(0.014,0.006)^{d}$  & \TphotSzero{} \\
Photometric GP frequency [$d^{-1}$] & $\mathcal{I}(3.525,0.651)^{d}$  &   \Tphotwzero{} \\
Photometric GP mean [ppt] & $\mathcal{I}(0.008,0.036)^{d}$  & \Tphotmean{} \\
\hline
\multicolumn{3}{l}{\harps{} parameters}\\
log RV semi-amplitude, $\log{K}$ b &  $\mathcal{N}(0.3,5.0)$  &   \TlogKzero{} \\
log RV semi-amplitude, $\log{K}$ c &  $\mathcal{N}(0.3,5.0)$  &   \TlogKone{} \\
RV trend - intercept at BJD=2458779.717 [\ms{}] &  $\mathcal{N}(0.0,0.1)$  & \Trvtrendzero{} \\
RV trend - gradient [\ms{}$d^{-1}$] &  $\mathcal{N}(0.0,1.0)$  &   \Trvtrendone{} \\
\harps{} log jitter RV [\ms{}] &  $\mathcal{N}(1.992,5.0)$  &   \Trvlogerrcontrzero{} \\
\harps{} log jitter S index &  $\mathcal{N}(6.527e-06,5.0)$  &  \Trvlogerrcontrone{} \\
\harps{} log jitter FWHM [\ms{}] &  $\mathcal{N}(20.525,5.0)$  &   \Trvlogerrcontrtwo{} \\
\harps{} mean S-index &  $\mathcal{N}(0.0,0.00941)$  &  \Tmeans{} \\
\harps{} mean FWHM [\ms{}] &  $\mathcal{N}(7287.75,7.5)$  &  \Tmeanfwhm{} \\
\harps{} GP log amplitude RV &  $\mathcal{N}(2.984,8.0)$  &  \Tlogamprv{} \\
\harps{} GP log amplitude S-index &  $\mathcal{N}(-9.332,8.0)$  & \Tlogamps{} \\
\harps{} GP log amplitude FWHM &  $\mathcal{N}(4.03,8.0)$  & \Tlogampfwhm{}  \\
\harps{} GP log rotation period, $\log{P_{\rm rot}}/\log{d}$ &  $\mathcal{N}_{\mathcal{U}}(3.024,0.2,1.099,4.382)$  & \Tlogperiod{} \\
\harps{} GP log quality, $Q$ &  $\mathcal{N}(0.0,10.0)$  &  \TlogQzero{} \\
\harps{} GP log quality differential, $\Delta Q$ &  $\mathcal{N}(0.0,5.0)$  & \TdeltaQ{}{} \\
\harps{} GP $P_{\rm rot}$ - $P_{\rm rot}/2$ mix factor &  $\mathcal{U}(0,1)$  & \Tperiod{} \\
\hline
\hline
\multicolumn{3}{l}{
  \begin{minipage}{14cm}
    $\mathcal{N}(\mu;\sigma^{2})$ is a normal distribution with mean $\mu$ and width $\sigma^{2}$, $\mathcal{U}(a;b)$ is a uniform distribution between $a$ and $b$, $\mathcal{N}_{\mathcal{U}}(\mu;\sigma^{2},a,b)$ is a normal distribution with mean $\mu$ and width $\sigma^{2}$ multiplied with a uniform distribution between $a$ and $b$, $\beta(a;b)$ is a Beta distribution with parameters $a$ and $b$, and $\mathcal{I}(\mu;\sigma^2)$ is a distribution directly interpolated from the output of a pre-trained distribution with mean $\mu$ and standard deviation $\sigma^2$ (although the distribution may not follow a normal distribution).  Posterior values and uncertainties represent the median and $1\sigma$ error boundaries. All other values (e.g. presented in Table \ref{tab:derived_pars}) are directly determined from these fitted quantities. The prior uncertainties of input parameters $t_0$ and $P$ were inflated from the input data uncertainties by factors of: $t_{0,b} = 23 \times$, $t_{0,c} = 7 \times$, $P_b = 3 \times$, $P_c = 11 \times$. $^{a}$Described in \citet{kipping2013parametrizing}.  $^{b}$Reparameterised in \texttt{exoplanet} to avoid discontinuities at $\pm\pi$. $^{c}$\texttt{exoplanet} reparameterization of \citet{espinoza2018efficient}. $^{d}$\texttt{PyMc3} Interpolation function of pre-trained GP.  \end{minipage}}
\end{tabular}
\end{center}
\label{AllPriors}
\end{table*}%

\begin{table}
\caption{\harps{} spectroscopy from first season (June - August 2019)}
\label{tab:spec1}
\scriptsize
\begin{tabular}{lcccccc}
\hline
\hline
Time & RV & $\sigma_{\rm RV}$ & $S_{\rm MW}$ & $\sigma_{S}$ & FWHM & $\sigma_{\rm FWHM}$ \\
$[\rm{BJD}-2457000]$ & \multicolumn{2}{c}{[\ms{}]} & \multicolumn{2}{c}{--} & \multicolumn{2}{c}{[\ms{}]} \\
\hline
\hline
$1655.5493$ & $1.8$ & $1.95$ & $0.0091$ & $0.004$ & $7281.8$ & $4.5$ \\
$1655.6181$ & $1.32$ & $2.0$ & $0.0003$ & $0.0045$ & $7291.5$ & $4.5$ \\
$1656.6167$ & $-4.08$ & $2.39$ & $0.0059$ & $0.0064$ & $7287.6$ & $4.5$ \\
$1657.5254$ & $-6.07$ & $1.44$ & $0.0042$ & $0.0025$ & $7277.2$ & $4.5$ \\
$1657.606$ & $-6.37$ & $1.49$ & $0.011$ & $0.0029$ & $7287.2$ & $4.6$ \\
$1658.5953$ & $-3.78$ & $1.46$ & $-0.0045$ & $0.0031$ & $7292.7$ & $4.6$ \\
$1661.5662$ & $0.04$ & $1.48$ & $-0.0048$ & $0.003$ & $7283.7$ & $4.5$ \\
$1664.5324$ & $-13.5$ & $1.36$ & $-0.0074$ & $0.0028$ & $7280.8$ & $4.5$ \\
$1664.6282$ & $-16.17$ & $1.6$ & $-0.0177$ & $0.004$ & $7276.6$ & $4.5$ \\
$1666.5674$ & $-6.22$ & $1.45$ & $-0.0164$ & $0.0029$ & $7284.6$ & $4.5$ \\
$1667.5542$ & $-8.01$ & $1.39$ & $-0.0127$ & $0.0028$ & $7285.4$ & $4.6$ \\
$1668.5189$ & $3.21$ & $1.8$ & $-0.0126$ & $0.0036$ & $7284.8$ & $4.5$ \\
$1668.6197$ & $5.3$ & $1.88$ & $-0.0178$ & $0.0044$ & $7289.2$ & $4.5$ \\
$1669.466$ & $-3.47$ & $1.32$ & $-0.0046$ & $0.0021$ & $7276.4$ & $4.5$ \\
$1669.5709$ & $-3.81$ & $1.42$ & $-0.015$ & $0.0026$ & $7277.4$ & $4.5$ \\
$1670.4637$ & $-3.35$ & $1.13$ & $-0.0009$ & $0.0015$ & $7284.6$ & $4.5$ \\
$1670.5832$ & $-3.13$ & $1.46$ & $-0.0052$ & $0.0028$ & $7293.5$ & $4.6$ \\
$1673.5985$ & $2.86$ & $1.53$ & $0.0004$ & $0.0032$ & $7292.2$ & $4.6$ \\
$1674.5613$ & $-0.04$ & $1.69$ & $-0.0008$ & $0.0038$ & $7296.6$ & $4.6$ \\
$1676.4716$ & $2.3$ & $1.62$ & $0.0035$ & $0.0034$ & $7302.6$ & $4.6$ \\
$1676.5886$ & $-2.22$ & $1.88$ & $-0.0014$ & $0.0051$ & $7297.6$ & $4.6$ \\
$1677.4681$ & $-5.59$ & $1.43$ & $0.0078$ & $0.0029$ & $7294.7$ & $4.6$ \\
$1677.5491$ & $-9.0$ & $1.68$ & $-0.0021$ & $0.0043$ & $7288.7$ & $4.6$ \\
$1679.5086$ & $-12.66$ & $1.12$ & $-0.001$ & $0.0019$ & $7282.5$ & $4.6$ \\
$1679.571$ & $-11.58$ & $1.56$ & $-0.0098$ & $0.0036$ & $7286.5$ & $4.6$ \\
$1680.5087$ & $-10.82$ & $1.09$ & $-0.0003$ & $0.0017$ & $7280.1$ & $4.6$ \\
$1680.5634$ & $-10.88$ & $1.16$ & $-0.0066$ & $0.0021$ & $7274.6$ & $4.6$ \\
$1681.5245$ & $-4.52$ & $1.54$ & $-0.0098$ & $0.0035$ & $7277.2$ & $4.5$ \\
$1681.5771$ & $-6.48$ & $1.75$ & $-0.0162$ & $0.0043$ & $7277.4$ & $4.5$ \\
$1682.4813$ & $-5.11$ & $1.23$ & $-0.0081$ & $0.002$ & $7278.9$ & $4.5$ \\
$1682.5533$ & $-9.81$ & $1.35$ & $-0.016$ & $0.0028$ & $7279.4$ & $4.6$ \\
$1684.5367$ & $-7.71$ & $1.75$ & $-0.0143$ & $0.0037$ & $7282.2$ & $4.5$ \\
$1684.5971$ & $-7.15$ & $1.49$ & $-0.0165$ & $0.0031$ & $7276.2$ & $4.5$ \\
$1685.4972$ & $-3.64$ & $1.39$ & $-0.0081$ & $0.003$ & $7279.2$ & $4.5$ \\
$1685.5436$ & $-7.43$ & $1.69$ & $-0.0174$ & $0.0044$ & $7281.9$ & $4.6$ \\
$1689.5056$ & $6.68$ & $1.47$ & $-0.0023$ & $0.0032$ & $7287.8$ & $4.6$ \\
$1689.5493$ & $5.35$ & $1.83$ & $-0.019$ & $0.0047$ & $7290.2$ & $4.6$ \\
$1690.4858$ & $4.11$ & $1.6$ & $-0.0033$ & $0.0032$ & $7280.8$ & $4.5$ \\
$1691.5335$ & $1.79$ & $1.58$ & $-0.005$ & $0.0037$ & $7287.9$ & $4.6$ \\
$1691.5549$ & $1.87$ & $1.62$ & $-0.0057$ & $0.0039$ & $7283.2$ & $4.6$ \\
$1692.5178$ & $4.23$ & $1.59$ & $-0.0053$ & $0.0032$ & $7294.8$ & $4.5$ \\
$1693.4664$ & $7.52$ & $1.36$ & $0.002$ & $0.0026$ & $7291.0$ & $4.5$ \\
$1694.4709$ & $8.59$ & $1.19$ & $0.0034$ & $0.002$ & $7288.2$ & $4.5$ \\
$1695.462$ & $6.68$ & $1.21$ & $0.0056$ & $0.0018$ & $7287.3$ & $4.5$ \\
$1697.4761$ & $0.74$ & $2.07$ & $-0.0023$ & $0.0052$ & $7288.8$ & $4.5$ \\
$1698.4702$ & $-4.52$ & $1.57$ & $0.0019$ & $0.0029$ & $7288.1$ & $4.5$ \\
$1699.4797$ & $-6.82$ & $1.3$ & $-0.0051$ & $0.0024$ & $7283.2$ & $4.5$ \\
$1700.4668$ & $-7.25$ & $1.29$ & $0.0007$ & $0.0023$ & $7284.8$ & $4.5$ \\
$1701.4669$ & $-2.83$ & $1.19$ & $-0.0044$ & $0.0021$ & $7272.8$ & $4.5$ \\
$1702.4715$ & $-5.04$ & $1.47$ & $-0.0185$ & $0.0032$ & $7275.2$ & $4.5$ \\
$1703.4744$ & $-13.62$ & $1.76$ & $-0.0192$ & $0.0042$ & $7269.0$ & $4.5$ \\
$1704.4713$ & $-10.69$ & $2.09$ & $-0.024$ & $0.0054$ & $7274.8$ & $4.5$ \\
$1705.4964$ & $-11.55$ & $1.38$ & $-0.0108$ & $0.0029$ & $7277.4$ & $4.5$ \\
$1706.4982$ & $-8.22$ & $1.47$ & $-0.0119$ & $0.0033$ & $7279.4$ & $4.5$ \\
$1707.5135$ & $-7.27$ & $2.37$ & $-0.0333$ & $0.0066$ & $7278.0$ & $4.5$ \\
$1708.4678$ & $0.39$ & $1.35$ & $-0.0052$ & $0.0026$ & $7286.6$ & $4.5$ \\
\hline
\hline
\end{tabular}
\end{table}
\begin{table}
\caption{\harps{} spectroscopy from second season (Dec 2019 - Feb 2020).}
\label{tab:spec2}
\scriptsize
\begin{tabular}{lcccccc}
\hline
\hline
Time & RV & $\sigma_{\rm RV}$ & $S_{\rm MW}$ & $\sigma_{S}$ & FWHM & $\sigma_{\rm FWHM}$ \\
$[\rm{BJD}-2457000]$ & \multicolumn{2}{c}{[\ms{}]} & \multicolumn{2}{c}{--} & \multicolumn{2}{c}{[\ms{}]} \\
\hline
\hline
$1838.8494$ & $11.17$ & $1.4$ & $0.0147$ & $0.0022$ & $7301.0$ & $4.5$ \\
$1839.8578$ & $6.52$ & $1.26$ & $0.0142$ & $0.0017$ & $7301.5$ & $4.5$ \\
$1840.8432$ & $3.27$ & $1.15$ & $0.0136$ & $0.0014$ & $7294.8$ & $4.5$ \\
$1841.8384$ & $3.16$ & $1.19$ & $0.0097$ & $0.0015$ & $7292.3$ & $4.5$ \\
$1842.8077$ & $-3.96$ & $1.45$ & $0.0039$ & $0.0022$ & $7286.0$ & $4.5$ \\
$1843.8559$ & $-2.03$ & $1.74$ & $-0.0018$ & $0.0029$ & $7284.0$ & $4.5$ \\
$1844.8362$ & $-13.35$ & $1.16$ & $-0.0026$ & $0.0014$ & $7274.8$ & $4.5$ \\
$1845.8271$ & $-12.38$ & $1.33$ & $-0.0073$ & $0.002$ & $7272.5$ & $4.5$ \\
$1847.8389$ & $-14.59$ & $1.11$ & $-0.0095$ & $0.0012$ & $7274.5$ & $4.5$ \\
$1849.7822$ & $-1.97$ & $1.18$ & $-0.0042$ & $0.0015$ & $7282.3$ & $4.5$ \\
$1849.8576$ & $-3.28$ & $1.13$ & $-0.0055$ & $0.0013$ & $7282.4$ & $4.5$ \\
$1850.7988$ & $0.75$ & $1.08$ & $0.0006$ & $0.0013$ & $7283.3$ & $4.5$ \\
$1850.8597$ & $2.51$ & $1.04$ & $-0.001$ & $0.0011$ & $7285.2$ & $4.5$ \\
$1852.7908$ & $-1.78$ & $1.17$ & $0.0036$ & $0.0015$ & $7289.0$ & $4.5$ \\
$1852.8609$ & $-0.05$ & $1.42$ & $0.0059$ & $0.0021$ & $7287.6$ & $4.5$ \\
$1853.8015$ & $1.87$ & $1.25$ & $0.0071$ & $0.0017$ & $7296.4$ & $4.5$ \\
$1853.8644$ & $6.84$ & $1.68$ & $0.0055$ & $0.003$ & $7298.1$ & $4.5$ \\
$1854.757$ & $4.48$ & $1.88$ & $-0.0022$ & $0.0043$ & $7301.3$ & $4.5$ \\
$1854.8269$ & $-1.07$ & $1.62$ & $-0.0036$ & $0.0031$ & $7287.3$ & $4.5$ \\
$1855.8321$ & $2.95$ & $1.38$ & $0.0034$ & $0.0022$ & $7291.6$ & $4.5$ \\
$1858.775$ & $7.83$ & $1.33$ & $0.0067$ & $0.002$ & $7300.9$ & $4.5$ \\
$1858.8351$ & $8.32$ & $1.31$ & $0.0079$ & $0.0018$ & $7296.7$ & $4.5$ \\
$1859.7796$ & $5.81$ & $1.44$ & $0.0076$ & $0.0024$ & $7297.4$ & $4.5$ \\
$1860.7517$ & $-2.5$ & $1.26$ & $0.008$ & $0.0018$ & $7289.7$ & $4.5$ \\
$1860.8525$ & $-4.61$ & $1.19$ & $0.0083$ & $0.0015$ & $7290.3$ & $4.5$ \\
$1861.7734$ & $2.82$ & $1.22$ & $0.0061$ & $0.0016$ & $7285.7$ & $4.5$ \\
$1861.85$ & $3.7$ & $1.22$ & $0.0075$ & $0.0015$ & $7289.9$ & $4.5$ \\
$1862.7588$ & $-0.94$ & $1.1$ & $0.0043$ & $0.0012$ & $7293.8$ & $4.5$ \\
$1862.8378$ & $-0.75$ & $1.19$ & $0.0089$ & $0.0014$ & $7296.8$ & $4.5$ \\
$1863.7554$ & $2.21$ & $1.13$ & $0.0048$ & $0.0014$ & $7294.8$ & $4.5$ \\
$1863.8328$ & $1.3$ & $1.06$ & $0.0039$ & $0.0011$ & $7292.6$ & $4.5$ \\
$1864.7737$ & $-0.21$ & $1.25$ & $0.0048$ & $0.0017$ & $7293.5$ & $4.5$ \\
$1864.8432$ & $-3.66$ & $1.46$ & $0.005$ & $0.0022$ & $7296.3$ & $4.5$ \\
$1865.8255$ & $-9.78$ & $1.41$ & $0.0104$ & $0.002$ & $7289.3$ & $4.5$ \\
$1876.7398$ & $6.07$ & $1.26$ & $0.0029$ & $0.0018$ & $7281.5$ & $4.5$ \\
$1876.8589$ & $7.09$ & $1.32$ & $-0.0009$ & $0.0018$ & $7277.0$ & $4.5$ \\
$1877.7559$ & $4.9$ & $1.25$ & $0.0009$ & $0.0017$ & $7285.7$ & $4.5$ \\
$1879.7785$ & $4.46$ & $1.56$ & $0.0045$ & $0.0025$ & $7290.9$ & $4.5$ \\
$1880.7355$ & $4.05$ & $1.22$ & $0.0087$ & $0.0016$ & $7294.9$ & $4.5$ \\
$1880.8847$ & $1.67$ & $1.22$ & $0.0083$ & $0.0017$ & $7296.2$ & $4.5$ \\
$1881.7267$ & $10.47$ & $1.23$ & $0.0099$ & $0.0016$ & $7289.8$ & $4.5$ \\
$1882.8435$ & $14.18$ & $1.23$ & $0.0126$ & $0.0015$ & $7295.8$ & $4.5$ \\
$1883.7334$ & $14.27$ & $1.17$ & $0.0151$ & $0.0015$ & $7300.2$ & $4.5$ \\
$1883.8645$ & $16.62$ & $1.26$ & $0.0136$ & $0.0017$ & $7302.6$ & $4.5$ \\
$1894.7258$ & $-3.86$ & $1.43$ & $-0.0031$ & $0.0021$ & $7283.7$ & $4.5$ \\
$1894.859$ & $-4.95$ & $1.49$ & $-0.0048$ & $0.0022$ & $7286.4$ & $4.5$ \\
$1897.8044$ & $8.9$ & $1.29$ & $-0.0005$ & $0.0017$ & $7291.1$ & $4.5$ \\
$1897.8921$ & $6.79$ & $1.5$ & $-0.0025$ & $0.0026$ & $7283.0$ & $4.5$ \\
$1898.8055$ & $8.8$ & $1.18$ & $0.0065$ & $0.0014$ & $7290.6$ & $4.5$ \\
$1899.7514$ & $10.63$ & $1.17$ & $0.0079$ & $0.0015$ & $7288.3$ & $4.5$ \\
$1899.8854$ & $7.06$ & $1.2$ & $0.0057$ & $0.0019$ & $7291.0$ & $4.5$ \\
$1900.7715$ & $3.46$ & $1.11$ & $0.0071$ & $0.0013$ & $7295.4$ & $4.5$ \\
$1900.8838$ & $3.59$ & $1.21$ & $0.0054$ & $0.002$ & $7289.1$ & $4.5$ \\
$1901.7655$ & $5.17$ & $1.07$ & $0.0084$ & $0.0012$ & $7289.1$ & $4.5$ \\
$1902.6953$ & $12.1$ & $1.13$ & $0.0102$ & $0.0014$ & $7289.0$ & $4.5$ \\
$1902.8507$ & $12.72$ & $1.16$ & $0.0106$ & $0.0017$ & $7291.3$ & $4.5$ \\
$1903.7072$ & $14.72$ & $1.07$ & $0.011$ & $0.0012$ & $7285.1$ & $4.5$ \\
$1903.885$ & $15.9$ & $1.26$ & $0.0071$ & $0.0022$ & $7290.0$ & $4.5$ \\
\hline
\hline
\end{tabular}
\end{table}


\bsp	
\label{lastpage}
\end{document}